# The size, shape, density and ring of the dwarf planet Haumea from a stellar occultation

**J. L. Ortiz[1], P. Santos-Sanz[1], B. Sicardy[2], G. Benedetti-Rossi[3], D. Bérard[2], N. Morales[1], R. Duffard[1], F. Braga-Ribas[3,4], U. Hopp[5,6], C. Ries[5], V. Nascimbeni[7,8], F. Marzari[9], V. Granata[7,8], A. Pál[10], C. Kiss[10], T. Pribulla[11], R. Komžík[11], K. Hornoch[12], P. Pravec[12], P. Bacci[13], M. Maestripieri[13], L. Nerli[13], L. Mazzei[13], M. Bachini[14,15], F. Martinelli[15], G. Succi[15], F. Ciabattari[16], H. Mikuz[17], A. Carbognani[18], B. Gaehrken[19], S. Mottola[20], S. Hellmich[20], F. L. Rommel[4], E. Fernández-Valenzuela[1], A. Campo Bagatin[21,22], S. Cikota[23,24], A. Cikota[25], J. Lecacheux[2], R. Vieira-Martins[3,26], J. I. B. Camargo[3,27], M. Assafin[28], F. Colas[26], R. Behrend[29], J. Desmars[2], E. Meza[2], A. Alvarez-Candal[3], W. Beisker[30], A. R. Gomes-Junior[28], B. E. Morgado[3], F. Roques[2], F. Vachier[26], J. Berthier[26], T. G. Mueller[6], J. M. Madiedo[31], O. Unsalan[32], E. Sonbas[33], N. Karaman[33], O. Erece[34], D. T. Koseoglu[34], T. Ozisik[34], S. Kalkan[35], Y. Guney[36], M. S. Niaei[37], O. Satir[37], C. Yesilyaprak[37,38], C. Puskullu[39], A. Kabas[39], O. Demircan[39], J. Alikakos[40], V. Charmandaris[40,41], G. Leto[42], J. Ohlert[43,44], J. M. Christille[18], R. Szakáts[10], A. Takácsné Farkas[10], E. Varga-Verebélyi[10], G. Marton[10], A. Marciniak[45], P. Bartczak[45], T. Santana-Ros[45], M. Butkiewicz-Bąk[45], G. Dudziński[45], V. Alí-Lagoa[6], K. Gazeas[46], L. Tzouganatos[46], N. Paschalis[47], V. Tsamis[48], A. Sánchez-Lavega[49], S. Pérez-Hoyos[49], R. Hueso[49], J. C. Guirado[50,51], V. Peris[50], R. Iglesias-Marzoa[52,53]**

(1) Instituto de Astrofísica de Andalucía (CSIC), Glorieta de la Astronomía s/n, 18008-Granada, Spain

(2) LESIA, Observatoire de Paris, PSL Research University, CNRS, Sorbonne Universités, UPMC Univ. Paris 06, Univ. Paris Diderot, Sorbonne Paris Cité, France

(3) Observatório Nacional/MCTIC, Rua Gal. José Cristino 77, Rio de Janeiro-RJ, CEP 20921-400, Brazil

(4) Federal University of Technology-Paraná (UTFPR / DAFIS), Rua Sete de Setembro, 3165, CEP 80230-901, Curitiba, PR, Brazil

(5) Universitäts-Sternwarte München, München, Scheiner Str. 1, D 81679 München, Germany

(6) Max-Planck-Institut für Extraterrestrische Physik, D 85741 Garching, Germany

(7) Dipartimento di Fisica e Astronomia, `G. Galilei', Università degli Studi di Padova, Vicolo dell'Osservatorio 3, I-35122 Padova, Italy

(8) INAF - Osservatorio Astronomico di Padova, vicolo dell'Osservatorio 5, I-35122 Padova, Italy

(9) Dipartimento di Fisica, University of Padova, via Marzolo 8, 35131, Padova, Italy

(10) Konkoly Observatory, Research Centre for Astronomy and Earth Sciences, Hungarian Academy of Sciences, Konkoly Thege 15-17, H-1121 Budapest, Hungary

(11) Astronomical Institute, Slovak Academy of Sciences, 059 60 Tatranská Lomnica, Slovakia

(12) Astronomical Institute, Academy of Sciences of the Czech Republic, Fričova 298, 251 65 Ondřejov, Czech Republic

(13) Astronomical Observatory San Marcello Pistoiese CARA Project, San Marcello Pistoiese, Pistoia, Italy

(14) Osservatorio astronomico di Tavolaia, Santa Maria a Monte (Pisa), Italy

(15) Lajatico Astronomical Centre, Via Mulini a Vento m. 9 Orciatico, cap 56030 Lajatico (Pisa), Italy

(16) Osservatorio Astronomico di Monte Agliale, Via Cune Motrone, I-55023 Borgo a Mozzano, Lucca, Italy

(17) Črni Vrh Observatory, Predgriže 29A, 5274 Črni Vrh nad Idrijo, Slovenia

(18) Astronomical Observatory of the Autonomous Region of the Aosta Valley (OAVdA), Lignan 39, 11020 Nus (Aosta), Italy

(19) Bayerische Volkssternwarte München, Rosenheimer Str. 145h D-81671 München, Germany

(20) German Aerospace Center (DLR), Institute of Planetary Research, Rutherfordstraße 2, 12489 Berlin, Germany

(21) Departamento de Física, Ingeniería de Sistemas y Teoría de la Señal, Universidad de Alicante, PO Box 99, E-03080 Alicante, Spain

(22) Instituto Universitario de Física Aplicada a las Ciencias y la Tecnología, Universidad de Alicante, PO Box 99, E-03080 Alicante, Spain

(23) University of Zagreb, Faculty of Electrical Engineering and Computing, Department of Applied Physics, Unska 3, 10000 Zagreb, Croatia

(24) Ruđer Bošković Institute, Bijenička cesta 54, 10000 Zagreb, Croatia

(25) European Southern Observatory, Karl-Schwarzschild-Str. 2, 85748 Garching b. München, Germany

(26) IMCCE/Observatorie de Paris, Paris, France, 77, Avenue Denfert Rochereau, 75014, Paris, France

(27) Laboratório Interinstitucional de e-Astronomia-LIneA, Rua Gal. José Cristino 77, Rio de Janeiro-RJ, CEP 20921-400, Brazil

(28) Observatório do Valongo/UFRJ, Ladeira Pedro Antônio 43, Rio de Janeiro-RJ, CEP 20080-090, Brazil

(29) Observatoire de Genève, CH1290 Sauverny, Switzerland

(30) International Occultation Timing Association - European Section (IOTA-ES) IOTA-ES e.V. Bartold-Knausstr. 8 D-30459 Hannover, Germany

(31) Facultad de Ciencias Experimentales, Universidad de Huelva, Avenida de las Fuerzas Armadas S/N. 21071 Huelva, Spain

(32) Ege University, Faculty of Science, Department of Physics, 35100, Bornova, Izmir, Turkey

(33) University of Adiyaman, Department of Physics, 02040 Adiyaman, Turkey

(34) TUBITAK National Observatory (TUG), Akdeniz University Campus, 07058, Antalya, Turkey

(35) Ondokuz Mayis University Observatory, Space Research Center, 55200, Kurupelit, Samsun, Turkey

(36) Atatürk University, Science Faculty, Department of Physics, 25240, Erzurum, Turkey

(37) Atatürk University, Astrophysics Research and Application Center (ATASAM), 25240, Erzurum, Turkey

(38) Atatürk University, Science Faculty, Department of Astronomy and Astrophysics, 25240, Erzurum, Turkey

(39) Canakkale Onsekiz Mart University, Astrophysics Research Center (ARC) and Ulupınar Observatory (UPO) Canakkale, Turkey

(40) Institute for Astronomy, Astrophysics, Space Applications & Remote Sensing, National Observatory of Athens, GR-15236, Penteli, Greece

(41) Department of Physics, University of Crete, GR-71003, Heraklion, Greece

(42) INAF-Catania Astrophysical Observatory, Via Santa Sofia 78, I-95123 Catania, Italy

(43) Michael Adrian Observatorium, Astronomie Stiftung Trebur, Fichtenstr. 7 65468 Trebur, Germany

(44) University of Applied Sciences, Technische Hochschule Mittelhessen, Wilhelm-Leuschner-Straße 13 D-61169 Friedberg, Germany

(45) Astronomical Observatory Institute, Faculty of Physics, A. Mickiewicz University, Słoneczna 36, 60-286 Poznań, Poland

(46) Section of Astrophysics, Astronomy and Mechanics, Department of Physics, National and Kapodistrian University of Athens, GR-15784 Zografos, Athens, Greece

(47) Nunki Observatory, Skiathos Island, Xanemo area 37002, Greece

(48) Ellinogermaniki Agogi Observatory, Dimitriou Panagea str., GR-15351 Pallini, Athens, Greece

(49) Departamento de Física Aplicada I, Escuela de Ingeniería de Bilbao, Universidad del País Vasco UPV /EHU, Plaza Torres Quevedo 1, 48013 Bilbao, Spain

(50) Observatori Astronòmic de la Universitat de València. Catedrático José Beltrán, 2. 46980, Paterna (Valencia), Spain

(51) Departament d'Astronomia i Astrofísica, Universitat de València, C. Dr. Moliner 50, E-46100 Burjassot, València, Spain

(52) Centro de Estudios de Física del Cosmos de Aragón, Plaza de San Juan 1, 2ª planta, 44001, Teruel, Spain

(53) Dpto de Astrofísica, Universidad de La Laguna, Avenida Astrofísico Fco Sánchez, s/n, 38200, La Laguna, Tenerife, Spain

**Among the four known transneptunian dwarf planets, Haumea is an exotic, very elongated and fast rotating body[1,2,3]. In contrast to the other dwarf planets[4,5,6], its size, shape, albedo and density are not well constrained. Here we report results of a multi-chord stellar occultation, observed on 2017 January 21. Secondary events observed around the main body are consistent with the presence of a ring of opacity 0.5, width 70 km and radius $2,287^{+75}_{-45}$ km. The Centaur Chariklo was the first body other than a giant planet to show a ring system[7], and the Centaur Chiron was later found to possess something similar to Chariklo's rings[8,9]. Haumea is the first body outside the Centaur population with a ring. The ring is coplanar with both Haumea's equator and the orbit of its satellite Hi'iaka. Its radius places it close to the 3:1 mean motion resonance with Haumea's spin period. The occultation by the main body provides an instantaneous elliptical limb with axes 1,704 ± 4 km × 1,138 ± 26 km. Combined with rotational light curves, it constrains Haumea's 3D orientation and its triaxial shape, which is inconsistent with a homogeneous body in hydrostatic equilibrium. Haumea's largest axis is at least 2,322 ± 60 km, larger than thought before. This implies an upper limit of 1,885 ± 80 kg m$^{-3}$ for Haumea's density, smaller and less puzzling than previous estimations[1,10], and a geometric albedo of 0.51 ± 0.02, also smaller than previous estimations[11]. No global $N_2$ or $CH_4$ atmosphere with pressures larger than 15 and 50 nbar (3-$\sigma$ limits), respectively, is detected.**

Within our program of physical characterization of Transneptunian Objects (TNOs), we predicted an occultation of the star URAT1 533-182543 by the dwarf planet (136108) Haumea and arranged observations as explained in Methods. Positive occultation detections were obtained on 2017 January 21, from twelve telescopes at ten different observatories. The instruments and the main features of each station are listed in Table 1.

As detailed in Methods, see also Fig. 1, the light curves (the normalized flux from the star plus Haumea versus time) show deep drops caused by Haumea near the predicted time (~3:09 UT). Because the drops are abrupt, Haumea must lack a global Pluto-like atmosphere. An upper limit on the surface pressure of a Nitrogen or Methane-dominated atmosphere is determined to be 15 nbar and 50 nbar at 3$\sigma$-level, respectively.

Square-well fits to the drops provide the times of star disappearance and reappearance at each site. Those times define an occultation chord for each site, and an elliptical fit to the chord extremities provides the instantaneous limb of Haumea, with values 1,704 ± 4 km × 1,138 ± 26 km for the major and minor axes of the ellipse (Fig. 2). The position angle of the minor axis is -76.3°±1.2°.

In addition to the main occultation, there are brief dimmings prior and after the main event. These dips are consistently explained by a narrow and dense ring around Haumea that absorbed about 50% of the incoming stellar flux. It has reflectivity similar to those found around the Centaurs Chariklo[7] and Chiron[8,9] or around Uranus[12] and Neptune[13]. From an elliptical fit to the positions of those brief events (projected in the plane of the sky), we obtain an apparent semi-major axis $a'_{\rm ring} = 2{,}287^{+75}_{-45}$ km and apparent semi-minor axis $b'_{ring}$= 541 ± 15 km for the ring (the error bars stemming from the uncertainty in the timing of the events, see Fig. 3) The position angle of the ring apparent minor axis is $P_{\rm ring}$= -74.3°±1.3°, which coincides with that of Haumea's limb fit to within error bars, suggesting that the ring lies in Haumea's equatorial plane, see below.

Under the assumption that the ring ellipse seen in Fig. 3 corresponds to the projection of a circular ring, we derive a ring opening angle $B_{ring}$= $arcsin(b'_{ring}/a'_{ring})$= 13.8° ± 0.5° ($B_{ring}$=0° corresponding to an edge-on geometry). A ring particle orbiting a triaxial body with semi-axes $a>b>c$ rotating around the axis $c$ has its average angular momentum $H_c$ along $c$ conserved. Then, the state of least energy for a collisional, dissipative disk with $H_c$ constant is an equatorial ring. Thus, Haumea's equator should also be observed under the angle 13.8°. This is consistent with the high amplitude of its rotational light curve which requires low values of $B$ (ref. 1). The values of $P_{ring}$ *and* $B_{ring}$ provide two possible solutions for the ring pole, with J2000 equatorial coordinates $(\alpha_p, \delta_p)$ = (285.1° ± 0.5°,-10.6° ± 1.2°) (Solution 1) and $(\alpha_p, \delta_p)$ = (312.3° ± 0.3°,-18.6° ± 1.2°) (Solution 2). Solution 1 is preferred, because it is consistent with the long-term photometric behavior of Haumea, and because it is coincident, to within error bars, with the orbital pole position of Haumea's main satellite, Hi'iaka, $(\alpha_p, \delta_p)$ = (283.0° ± 0.2°, −10.6° ± 0.7°) (ref. 14). In that context, both the ring and Hi'iaka would lie in Haumea's equatorial plane.

At about 2,287 km from Haumea's center, the ring is within the Roche limit of a fluid-like satellite, which corresponds to ~4,400 km for a spherical Haumea (using a density of 1,885 kg/m$^3$ for the primary and a density of 500 kg/m$^3$ for the satellite). When the elongated shape of Haumea is considered, the Roche limit is even further out. Hence, the ring is close enough to Haumea so that accretion cannot proceed to form a satellite. The ring is close to the 3:1 spin-orbit resonance with Haumea, a ring particle undergoing one revolution while Haumea completes three rotations. This resonance occurs at 2,285 ± 8 km (see Methods). More knowledge of the ring orbit and Haumea's internal structure (which may be not homogeneous, see below) will be required to show if this proximity is just coincidental, or if the ring is actually trapped into this resonance and if it is, for what reason. Answering those questions, however, remains out of reach of the present work.

Another important property of Haumea is its geometric albedo ($p_V$), which can be determined using its projected area, as derived from the occultation, and its absolute magnitude[15]. We find a geometric albedo $p_V$= 0.51 ± 0.02, which is considerably smaller than the 0.7-0.75 and 0.804 $^{+0.062}_{-0.095}$ values derived from the latest combination of Herschel and Spitzer thermal measurements[16,8]. The geometric albedo should be even smaller if the contributions of the satellites and the ring to the absolute magnitude is larger than the 13.5% value used here (see Methods).

Since Haumea is thought to have a triaxial ellipsoid shape[1,17,18] with semi-axes $a>b>c$, the occultation alone cannot provide its three-dimensional shape unless we use additional information from the rotational light curve. From measurements performed days prior and after the occultation and given that we know Haumea's rotation period with high precision[16] we determined the rotational phase at the occultation time. It turns out that Haumea was at its absolute brightness minimum, which means that the projected area of the body was at its minimum during the occultation.

The magnitude change from minimum to maximum determined from the Hubble Space Telescope is 0.32 mag (using images that separated Haumea and Hi'iaka ref. 7), and using the equation 5 in ref. 19 together with the aspect angle in 2009 (when the observations were taken[7]) and the occultation ellipse parameters, we derive the following values for semi-axes of the ellipsoid, $a$=1,161 ± 30 km, $b$=852 ± 4 km and $c$= 513 ± 16 km (see Methods). The resulting density of Haumea, using its known mass[20] turns out to be 1,885 ± 80 kg/m$^3$, and its volume-equivalent diameter is 1,595 km ± 11 km. This is under the assumption that the ring does not contribute to the total brightness. For an upper limit of 5% contribution (see Methods), the real amplitude of the rotational light curve increases, and hence the a semiaxis increases too. The volume-equivalent diameter becomes 1,632 km and the density 1,757 kg/m$^3$. These two densities are significantly smaller than the lower limit of 2,600 kg/m$^3$ based on the figures of hydrostatic equilibrium, or based on mass and previous volume determinations[1]. A value in the range 1,885 to 1,757 kg/m$^3$ is far more in line with the density of other large TNOs and in agreement with the trend of increasing density versus size (e.g. supplementary material in ref. 5, and refs. 21, 22). We must also note that the axial ratios derived from the occultation are not consistent with those expected from the hydrostatic equilibrium figures of a homogeneous body[23] for the known rotation rate and the derived density. A previous work[24] had already hypothesized that the density of Haumea could be much smaller than the minimum 2,600 kg/m$^3$ value reported in the literature (if granular physics is used to model the shape of the body, instead of the simplifying assumption of fluid behavior). According to figure 4 of ref. 25 one can approximately determine an angle of friction between 10° and 15° for Haumea given the $c/a$ ratio of ~0.4 determined here. For reference, the maximum angle of friction of solid rocks on Earth is 45° and that of a fluid is 0°. Also, differentiation and other effects may play an important role in determining the final shape[26].

Chariklo, a body of around 250 km in diameter with a Centaur orbit (between the orbit of Jupiter and that of Neptune), was the first solar system object apart from the giant planets to show a ring system[7]. Shortly after that discovery, similar occultation features that resembled those from Chariklo's rings were found on Chiron[8,9], another Centaur. This had directed our attention to Centaurs and phenomenology related to them in order to explain these unexpected findings. The discovery of a ring around Haumea, a much more distant body, in a completely different dynamical class, much larger than Chariklo and Chiron, with satellites and with a very elongated triaxial shape, has numerous implications, such as rings being possibly common also in the transneptunian region from which Centaurs are delivered, and it opens the door to new avenues of research.

## ACKNOWLEDGEMENTS

These results were based on observations made with the 2m telescope at Wendelstein Observatory, which is operated by the Universitäts-Sternwarte München, the 1.8m telescope at Asiago Observatory, operated by Padova

Observatory, a member of the National Institute for Astrophysics, the 1.3m telescope at Skalnate Pleso Observatory, operated by the Astronomical Institute of the Slovak Academy of Science, the 1m telescope at Konkoly observatory, operated by Astrophysical Institute of the Hungarian Academy of Sciences, the 0.65m telescope at Ondrejov Observatory, operated by the Astronomical Institute of the Czech Academy of Sciences, the 1.5m telescope at Sierra Nevada Observatory, operated by the Instituto de Astrofisica de Andalucia-CSIC, the 1.23m telescope at Calar Alto Observatory, jointly operated by the Max Planck Institute für Astronomie and the IAA-CSIC, the Roque de los Muchachos Observatory 2m Liverpool telescope, operated by the Astrophysics Research Institute of Liverpool John Moores University, the Roque de los Muchachos Observatory 2.5m NOT telescope, operated by the Nordic Optical Telescope Scientific Association, the 1m telescope at Pic du Midi Observatory, operated by the Observatoire Midi Pyrénées, and the La Hita 0.77m telescope, which is jointly operated by Astrohita and the IAA-CSIC. JLO acknowledges funding from Spanish and Andalusian grants MINECO AYA-2014-56637-C2-1-P and J. A. 2012-FQM1776 as well as FEDER funds. Part of the research leading to these results received funding from the European Union's Horizon 2020 Research and Innovation Programme, under Grant Agreement no 687378. BS acknowledges support from the French grants "Beyond Neptune" ANR-08-BLAN-0177 and "Beyond Neptune II" ANR-11-IS56-0002. Part of the research leading to these results has received funding from the European Research Council under the European Community's H2020 (2014-2020/ERC Grant Agreement no. 669416). AP and RSz have been supported by the grant LP2012-31 of the Hungarian Academy of Sciences. GBR, FBR, FLR, RVM, JIBC, MA, ARGJ and BEM acknowledge the support from CAPES, CNPq and FAPERJ. JCG acknowledges funding from AYA2015-63939-C2-2-P and from the Generalitat Valenciana PROMETEOII/2014/057. KH and PP were supported by the project RVO:67985815. The Astronomical Observatory of the Autonomous Region of the Aosta Valley acknowledges a Shoemaker NEO Grant 2013 from The Planetary Society. Funds from a 2016 "Research and Education" grant from Fondazione CRT are acknowledged. The slovakian project ITMS No.26220120029 is also acknowledged.

## AUTHOR CONTRIBUTIONS

J.L.O. planned the campaign, analysed data for the prediction, made the prediction, participated in the observations, obtained and analysed data, interpreted the data and wrote the paper. P. S.-S. helped plan the campaign, analyzed data, helped interpret the data and helped write the paper. B.S. helped plan the campaign, analysed data, interpreted data, and wrote part of the paper. G. B.-R., and D. B. helped plan the campaign, participated in the observations, analysed and interpreted data. All other authors participated in the planning of the campaign and/or the observations and/or the interpretations. All authors were given the opportunity to review the results and comment on the manuscript.

## AUTHOR INFORMATION

The authors declare no competing financial interests. Correspondence and requests for materials should be addressed to JLO (ortiz@iaa.es).

## LEGENDS TO TABLES AND FIGURES

Table 1. Details of the observations on 21 January 2017. Observing sites from which the most relevant observations were obtained, main parameters of the observations for each site and the derived disappearance and reappearance times of the star caused by the central body, using the square well model fits in Fig. 1. The cycle times are the times between consecutive exposures. Note that at Wendelstein Observatory and at Konkoly Observatory two different telescopes were used. Hence there are 12 detections of the occultation from 10 different sites. Weather was clear in all the stations except at the Bavarian Public Observatory (Munich) where intermitent clouds were present.

Figure 1. Light curves of the occultation. Panels a) and b), light curves in the form of normalized flux versus time (at mid exposure), obtained from the different observatories that recorded the occultation (Table 1). The black dots and black lines represent the light curves extracted from the observations. The blue lines show the best square-well-model fits to the main body and the ring at Konkoly, with square-well models derived from the assumed ring width and opacity ($W$=70 km and $p'$=0.5) at other sites. The red dots and lines correspond to the optimal synthetic profile deduced from the square-well model fitted at each data point (see Methods). The rectangular profile in green line corresponds to the ring egress event at Skalnate Pleso, which fell in a readout time of the camera (see Fig. 3). Note that the light curves have been shifted in steps of 1 vertically for better viewing. Also note that “Munich” corresponds to the Bavarian Public Observatory. Error bars are 1σ.

Figure 2. Haumea’s projected shape. The blue lines are the occultation chords by the main body projected in the sky plane, as seen from nine observing sites (Table 1). The red segments are the uncertainties (1σ level) on each chord extremites, as derived from the timing uncertainties of Table 1. Note that we show the chord from Crni Vrh in dashed line because it is considerably uncertain. Note also that for the observatories where two telescopes were used we show only the best chord. The celestial North (“N”) and East (“E”) are indicated in the upper right

corner, together with the scale, and the arrow shows the star motion relative to the body. Haumea's limb (assumed elliptical) has been fitted to the chords, accounting for the uncertainties on each chord extremity (red segments). The limb has semi-major axis $a'$= 852±2 km and semi-minor axis $b'$=569±13 km, the latter having a position angle $P_{\rm limb}$= -76.3°±1.2° counted positively from the celestial north to celestial east. Haumea's equator has been drawn assuming that it is coplanar with the ring, with planetocentric elevation $B_{\rm ring}$= 13.7°±0.5°, see Fig. 3. The pink ellipse indicates the 1$\sigma$-level uncertainty domain for the ring center, while the blue ellipse inside it is the corresponding domain for Haumea's center. To within error bars, the ring and Haumea's centers (separated by 33 km in the sky plane) cannot be distinguished, showing so that our data are consistent with a circular ring concentric with the dwarf planet. The "a", "b" and "c" labels indicate the intersections of the a-, b- and c- semi-axes of the modeled ellipsoid with Haumea's surface.

Figure 3. Haumea's ring geometry. Fit to the ring events (red segments) with the same conventions as in Fig. 2. Those segments show the 1$\sigma$ uncertainty intervals for the midtimes of the secondary events at Mount Agliale (Ag), Lajatico (L), S. Marcello Pistoiese (SMP), Asiago (As), Wendelstein (W), Ondrejov (O), Konkoly (K), and Skalnate Pleso (S). Note that from the Bavarian Public Observatory (in Munich) no ring event could be detected because of the low signal to noise ratio. The ring egresses at Wendelstein, Asiago, S. Marcello Pistoiese and Lajatico are not observed because they are blocked by the main body. At Skalnate Pleso the ring egress is not detected (despite the high signal to noise ratio of the data) either because the ring is not homogeneous or because it is lost in the readout time (marked here in green). This is the most likely explanation because the readout times of 5.5s were long compared to the integration time of 10s. Note also that the green segment is very close to the positive Konkoly detection, making the hypothesis of an inhomogeneous ring unlikely. The two ellipses around Haumea delineate a 70-km wide ring with apparent opacity 0.5 (gray area) and semi-major axis $a_{\rm ring}$ = 2,287 $^{+75}_{-45}$ km that best fits simultaneously the secondary events of Fig. 1. The ring fit provides an opening angle $B_{\rm ring}$= 13.8°±0.5°, and a position angle for the ring apparent minor axis of $P_{\rm ring}$= -74.3°±1.3°. This is aligned, to within the error bars, with Haumea's apparent minor axis $P_{\rm limb}$=-76.3°±1.2° (Fig. 2). Moreover, Hi'iaka's orbital pole position[14] implies a sub-observer elevation $B_{\rm Hi'iaka}$=-15.7° above Hi'iaka's orbit on January 21, 2017,

and a superior conjunction occurring at position angle $P_{Hi'iaka}$= -73.6°. The fact that $|B_{ring}|$~$|B_{Hi'iaka}|$ and $P_{ring}$~$P_{limb}$~$P_{Hi'iaka}$ strongly suggests that both the ring and Hi'iaka orbit in Haumea's equatorial plane.

## TABLES AND FIGURES

Table 1.

| Site Name<br>Country and abbreviation | Coordinates<br>Lat dd:mm:ss<br>Lon dd:mm:ss<br>Altitude (m) | Telescope Aperture<br>Filter<br>Observer | Detector/Instrument<br>Exposure (s)<br>Cycle time (s) | Ingress time and<br>Egress time<br>(UTC) |
|---|---|---|---|---|
| Skalnate Pleso Observatory<br>-Slovakia (S) | 49°11'21.8'' N<br>20°14'02.1'' E<br>1826 | 1.3 m<br>no filter<br>R. Komžík | Moravian G4-9000<br>10 s<br>15.5 s | 3:08:26.79±0.96<br>3:10:24.56±0.8 |
| Konkoly Observatory<br>-Hungary<br>(K) | 47°55'01.6'' N<br>19°53'41.5'' E<br>935 | 1.0 m<br>no filter<br>A. Pál | Andor iXon-888<br>1 s<br>1.007 s | 3:08:20.3±0.2<br>3:10:17.39±0.07 |
| Konkoly Observatory<br>-Hungary<br>(K) | 47°55'01.6'' N<br>19°53'41.5'' E<br>935 | 0.6 m<br>no filter<br>A. Pál | Apogee Alta U16HC<br>2 s<br>2.944 s | 3:08:19.5±0.8<br>3:10:16.4±1.3 |
| Ondrejov Observatory<br>-Czech Republic<br>(O) | 49°54'32.6'' N<br>14°46'53.3'' E<br>526 | 0.65 m<br>no filter<br>K. Hornoch | Moravian G2-3200<br>8 s<br>9.721 s | 3:08:29.2±0.8<br>3:10:12.2±0.8 |
| Crni Vrh observatory<br>-Slovenia<br>(CV) | 45°56'45.0'' N<br>14°04'15.9'' E<br>713 | 0.6 m<br>clear<br>H. Mikuz | Apogee Alta U9000HC<br>300s, drifted<br>315s | 3:07:54±8<br>3:09:57±10 |
| Wendelstein Observatory<br>-Germany<br>(W) | 47°42'13.6'' N<br>12°00'44.0'' E<br>1838 | 2.0 m<br>r'<br>C. Ries | WWFI<br>10 s<br>14.536 s | 3:08:27.9±2.8<br>3:09:34.1±0.5 |
| Wendelstein Observatory<br>-Germany<br>(W) | 47°42'13.6'' N<br>12°00'44.0'' E<br>1838 | 0.4 m<br>r'<br>C. Ries | SBIG ST10-XME<br>30 s<br>53.096 s | 3:08:18.8 ±6<br>3:09:38.6 ±6 |
| Bavarian Public Observatory<br>-Munich, Germany (M) | 48°07'19.2'' N<br>11°36'25.2'' E<br>538 | 0.8m<br>no filter<br>B. Gaehrken | ATIK 314L+<br>20 s<br>20.304 s | 3:08:30.0±3.3<br>3:09:30.0±4.9 |
| Asiago observatory<br>Cima Ekar<br>-Italy (As) | 45°50'54.9"N<br>11°34'08.4"E<br>1376 | 1.82 m<br>no filter<br>V. Granata | AFOSC<br>2 s<br>5.026 s | 3:08:20.17±0.08<br>3:09:13.27±1.5 |
| S. Marcello Pistoiese observatory<br>-Italy<br>(SMP) | 44°03'51.0'' N<br>10°48'14.0'' E<br>965 | 0.6m<br>no filter<br>P. Bacci, M. Maestripieri, L. Nerli, L. Mazzei | Apogee Alta U6<br>10 s<br>11.877 s | 3:08:22.9±0.9<br>3:08:42.8±0.9 |
| Lajatico Astronomical Centre<br>-Italy<br>(L) | 43°25'44.7" N<br>10°43'01.2" E<br>433 | 0.5 m<br>no filter<br>M. Bachini, F. Martinelli, G. Succi | Moravian G3-1000<br>15 s<br>16.254 s | 3:08:19.9±1.4<br>3:08:34.3±1.4 |
| Mount Agliale observatory<br>-Italy (Ag) | 43°59’43.1’’ N<br>10°30’53.8’’ E<br>758 | 0.5 m<br>no filter<br>F. Ciabattari | FLI proline 4710<br>15 s<br>16.724 s | * |

*The occultation by the main body of Haumea was not detectable from Mount Agliale. Only ring events were detected from that observatory. See Methods and Extended Data Table 2.

Figure 1.

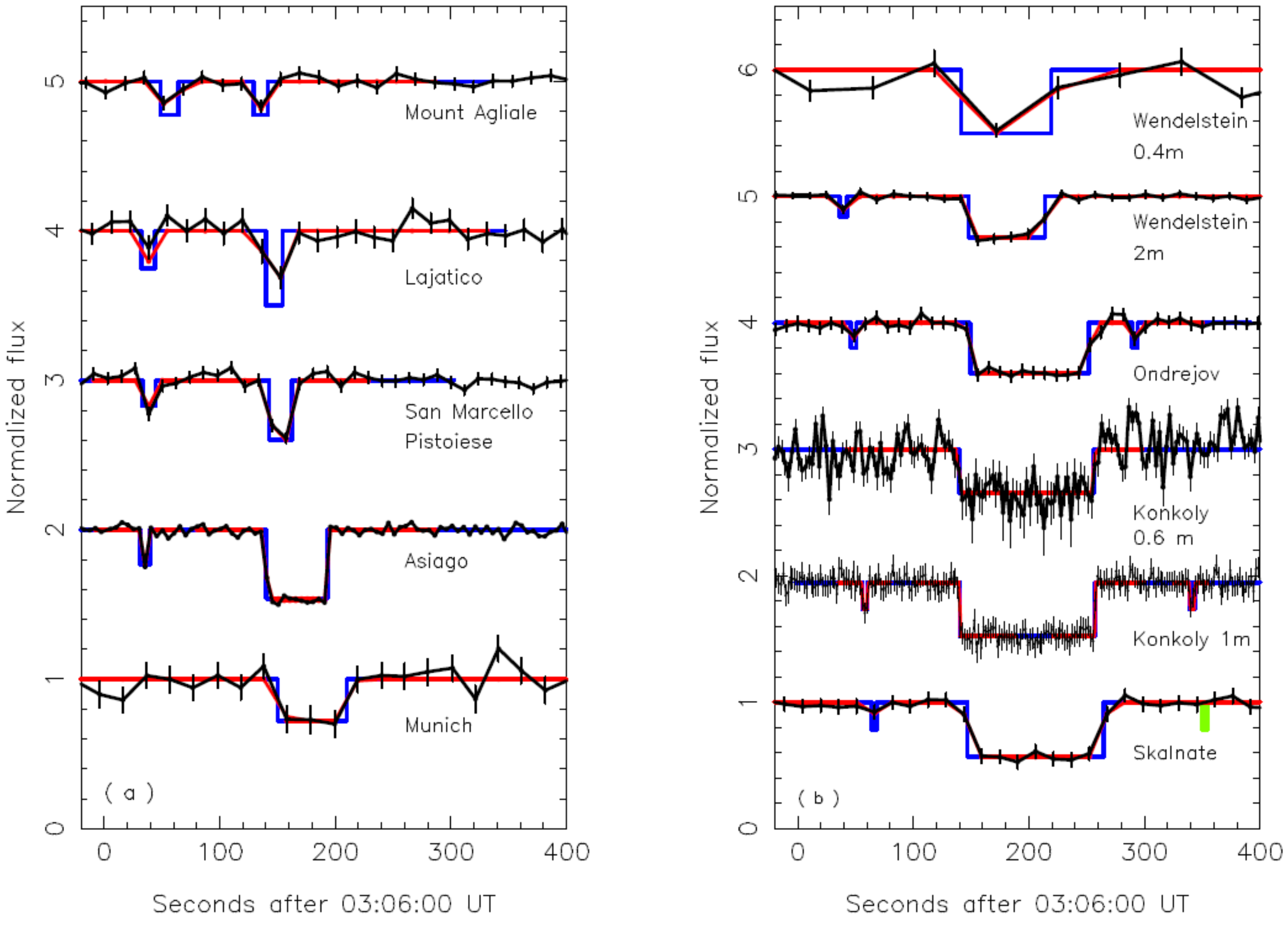

Figure 2.

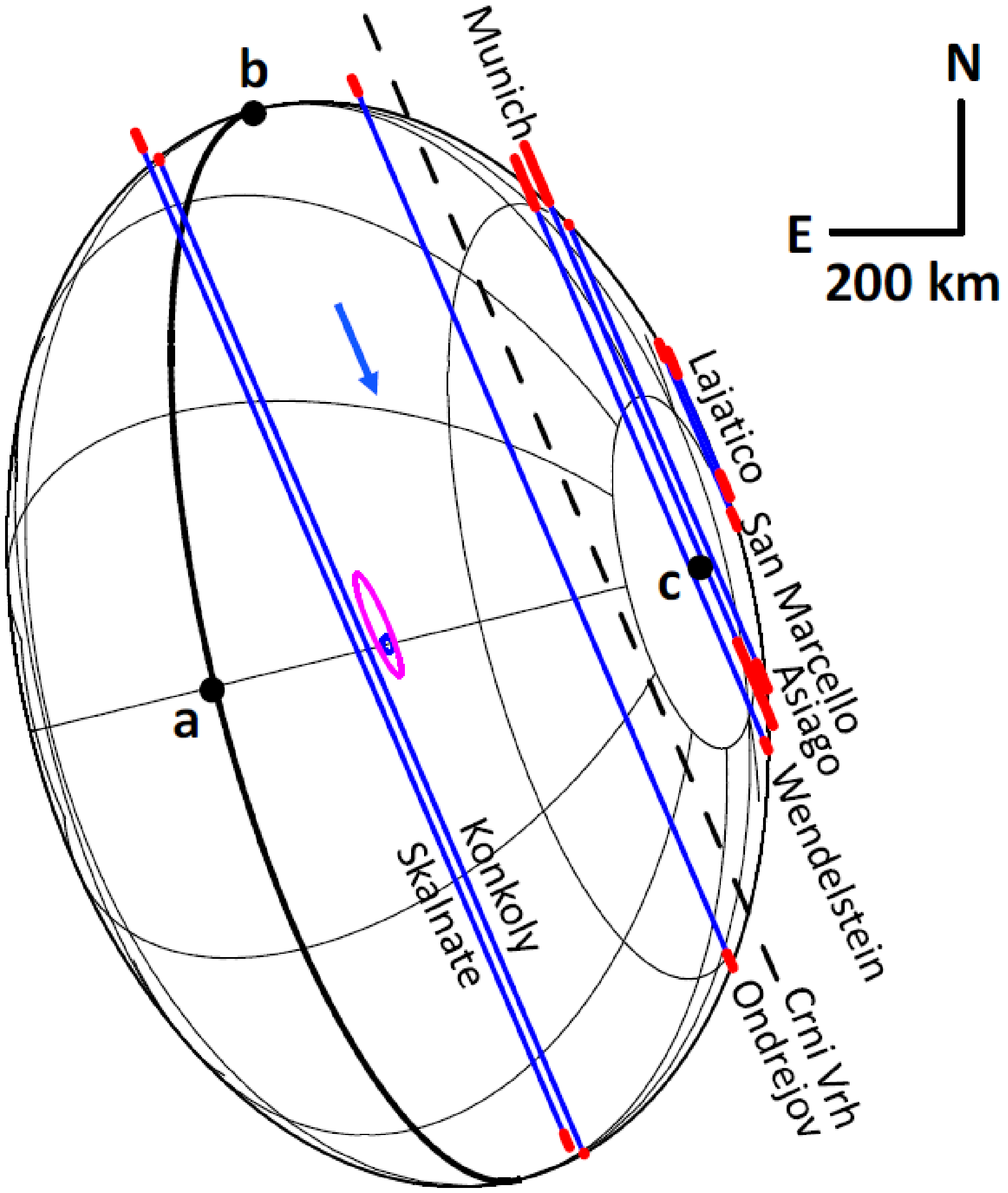

Figure 3.

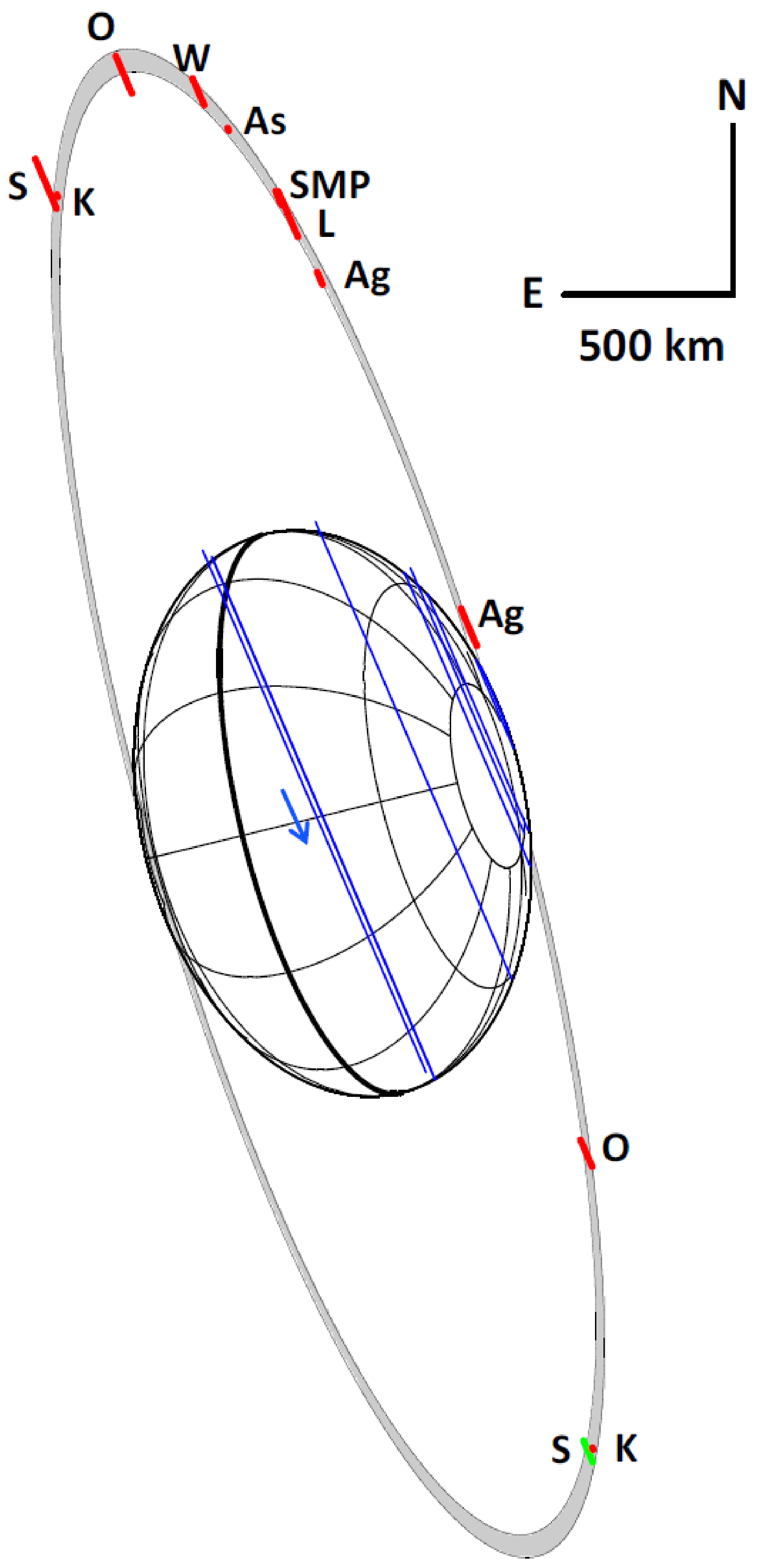

## METHODS

**OCCULTATION PREDICTIONS.** The occultation was predicted in late 2015 by using the URAT1 star catalog and several ephemeris sources for Haumea. Because the star involved in the occultation (URAT1 533-182543) was sufficiently bright ($m_R$~17.7), the event had the potential to be detectable from numerous sites. The initial predictions based on a JPL orbit (obtained at the JPL horizons web site[27]) put the shadow path slightly outside the Earth. Other ephemeris sources such as Astorb[28] and NIMA[29] were somewhat more favorable, giving rise to shadow paths well within the Earth. The scatter in the predictions, due to the uncertainty in Haumea's orbit, was of the order of 300 milliarcseconds (mas). Given that Haumea's angular diameter was expected to be much smaller than that, refined predictions were needed to make sure that a detection could be made. Besides, the presence of a large satellite (Hi'iaka) orbiting Haumea could cause an astrometric wobble similar to that detected in the system of Orcus and its satellite Vanth[30] due to the photocenter oscillation around the barycenter of the system.

Hence, we decided to carry out a detailed astrometric monitoring of Haumea. More than 1000 measurements of Haumea's position were obtained with the 0.77m f/3 La Hita telescope from April 1st to July 4th, 2016, using a 4k x 4k CCD (Charge-Coupled Device) camera. The unfiltered images were taken on a daily basis (weather and moon permitting) when Haumea was near culmination to avoid possible Differential Chromatic Refraction (DCR) problems as much as possible. The instrument, the image processing and the methodology were the same as those used for the prediction of the stellar occultation by Makemake[5]. The analysis of the data resulted in a shadow path favorable for locations in a wide region of the Earth. Hence, this encouraged further work. Once the GaiaDR1 catalog was released we improved our prediction with new coordinates of the occultation star by combining the position (epoch 2015.0) with the information on position and proper motions from the URAT1 at its 2013.669 and from other catalogs.

The final star coordinates (J2000) that we derived for the occultation epoch (2017.058) were: $\alpha$ = 14h 12m 03s.2034, $\delta$ = +16° 33' 58''.642, where $\alpha$ and $\delta$ are right ascension and declination respectively. After the occultation, the HSOY catalog[31] has become available, which provides coordinates and proper motions of GaiaDR1 stars matched with PPMXL stars. From this catalog, the J2000 coordinates for epoch 2017.058 turn out to be $\alpha$ = 14h 12m 03s.2034, $\delta$ = +16° 33' 58''.647, which completely agree in right ascension and differ by only 4 mas in declination.

Once the GaiaDR1 catalog became available in September 2016 we also redid the Haumea astrometry from the entire La Hita 0.77m telescope data set with respect to GaiaDR1. Thanks to the higher accuracy of GaiaDR1 with respect to previous catalogs of similar depth the resulting astrometry showed a clear oscillation of the residuals to the JPL orbit with respect to the measurements (Observed-Calculated) due to the presence of the satellite Hi'iaka. A Lomb periodogram[32] of the residuals in declination showed its strongest peak at a significant periodicity of 49.5 ± 0.9 days, which coincides with the known orbital period of Hi'iaka (49.462 ± 0.083 days according to ref. 20). A sinusoidal fit to the residuals (Extended Data Fig. 1) using the orbital period of Hi'iaka had a maximum when Hi'iaka's theoretical position was at its northernmost position with respect to Haumea, and the minimum of the fit corresponded to the southernmost position of the satellite Hi'iaka. Hence we verified that the oscillation was indeed correlated with the theoretical positions of Hi'iaka. For the theoretical computations of the Hi'iaka orbit we used the miriade ephemeris service[33]

With that information we could already make a reliable prediction of the occultation for 21 January 2017 that indicated a favorable shadow path for central Europe. The resulting offsets with respect to JPL#81 orbit (with planetary ephemeris version DE431) were 174 mas and 73 mas in $\alpha$*cos($\delta$) and in declination respectively. This represented the offset of the barycenter of Haumea's system with respect to the JPL#81 ephemerides. In early December 2016 we took again images of Haumea, and astrometrically solved them with respect to the GaiaDR1 catalog. This time we used several larger telescopes (the Sierra Nevada Observatory 1.5m telescope, the Calar Alto Observatory 1.2m telescope, the Pic du Midi Observatory 1m telescope, the 2m Liverpool telescope and the 2.5m Nordic Optical telescope to obtain observations of Haumea in R band). We determined the offsets of the astrometric measurements with respect to JPL ephemeris (Observed-Computed). Then we did a correlation analysis of the offsets with respect to the computed theoretical positions of the satellite Hi'iaka and in the end we came up with offsets in $\alpha$ and $\delta$ for the position of the barycenter of the Haumea system with respect to the ephemerides. These offsets were determined from linear fits and turned out to be 176 ± 6 mas in $\alpha$*cos($\delta$) and 73 ± 11 mas in $\delta$ with respect to JPL#81 orbit. The errors were determined from the uncertainties in the parameters of the linear fits. These offsets were in perfect agreement with those calculated with the large La Hita data set. The Spearman correlation coefficients were 0.91 and 0.88 for the right ascension and declination residuals respectively.

On the other hand, we determined the distance of Haumea with respect to the barycenter using the equation 1,232/(1+$r$) where 1,232 is the separation of Hi'iaka with respect to Haumea in mas and $r$ is the mass ratio of Haumea to Hi'iaka. That mass ratio was thought to be of the order 200, assuming similar albedos for Haumea and Hi'iaka (or even higher for Hi'iaka, given the depth of the water ice feature as shown in ref. 34). Projected in the plane of the sky, the offset position of Haumea on 21 January 2017 with respect to the barycenter was then calculated to be -6 mas in declination and 2 mas in $\alpha$*cos($\delta$). The predicted shadow path was favorable for Italy and central Europe. We decided to organize a campaign with observing sites from Spain to Turkey to maximize the chances of success and to compensate for possible unknown systematic errors that could have not been accounted for.

A map of the final shadow path on Earth with the sites that played the key roles in the observations is shown in Extended Data Fig. 2. The center of Haumea from the occultation was (17,16) mas away from the prediction, in α*cos(δ) and in δ, respectively. This translated into a difference of 409 km in the centerline of the shadow path and 55s in time with respect to the prediction.

**OCCULTATION OBSERVATIONS.** Sequences of CCD images were acquired at each observing site listed in Table 1. At Wendelstein, a red r' sloan filter was used, while all the other observatories used no filters to maximize the signal to noise ratio (SNR). Except for the Konkoly 1m-telescope observations, all the image sequences had interruptions due to the nonnegligible readout time of the cameras.

From the sequences of CCD images obtained at each telescope (and after standard dark-subtraction and flatfielding correction), light curves were constructed by carrying out synthetic aperture photometry of the occultation star (blended with Haumea) with respect to reference stars in the field of view. The synthetic aperture measurements were made using DAOPHOT routines and searching for the optimum aperture to minimize the scatter of the photometry. The photometry was performed relative to reference stars in the images so that small transparency fluctuations or seeing variations can be monitored and compensated for. The timing information was extracted from the time in the fits headers of the images. The time tagging accuracy at the different observatories (time tagging was done by synchronizing the controling computers with NTP time servers), is estimated to be 0.1s on average. The main parameters of the star are shown in Extended Data Table 1. It must be noted that the brightness of the star was similar to that of Haumea so at the time of the occultation we expected a brightness change of around 50% of the Haumea+star blended source. The resulting light curves (photon flux relative to the average value, versus time) are shown in Fig. 1.

The observations at Crni Vrh consisted of drifted images with 300s of exposure time, tracked at a speed of 40"/minute in the North-South direction. Hence aperture photometry could not be made. The detailed analysis procedure for this dataset is explained in a separate paragraph.

**LIMB FIT TO THE OCCULTATION CHORDS.** Square well fits to the occultation profiles were performed in order to accurately determine the times of star disappearance and reappearance for each observatory (Table 1) except for Crni Vrh, for which a special technique was needed as explained in a separate paragraph. The methodology to fit square wells was the same as that in other works of stellar occultations[4,5,7,35] by TNOs. The main parameters of the model are the depth of the square well and the disappearance time as well as the reappearance time of the occultation. The uncertainties in the retrieved parameters were obtained from a grid search in the parameter space. Acceptable values were those that gave a $\chi^2$ within $c^2_{\min}$ and $\chi^2_{\min}+1$. The uncertainties are listed in table 1. Note that the errors from the time tagging accuracy were an order of magnitude smaller than the uncertainties arising from the fits. From the fitted times at the different sites, one can generate chords in the projected plane of the sky. We fitted an elliptical limb to the extremities of the chords by minimizing a $\chi^2$ function defined as follows:

$$\chi^2 = \sum_1^N \frac{(r_{i,obs} - r_{i,cal})^2}{\sigma_{i,r}^2} \qquad (1)$$

where *r is* radius from the center of the ellipse *($f_c$,$g_c$),* the suffix *"cal"* means calculated and the suffix "*obs*" means observed. In this case $\sigma_{i,r}$ are the errors of the extremities, which were derived from the errors in the retrieved ingress and egress times (using the known speed of Haumea) and *N* is the number of extremities. The retrieved *a', b'* parameters (semi-axes of the fitted ellipse, not to be confused with the semiaxes of the triaxial ellipsoid, *a,b,c*) were translated from milliarcseconds into length in km by using the known distance of Haumea from Earth (50.4847 AU). This is the same procedure followed in other stellar occultation studies of TNOs[4,5,7]. The parameters of the ellipse fit were 1,704 ± 4 km, 1,138 ± 26 km, and $P_{\text{limb}}$=-76.3°±1.2°. The fit used 9 occulting chords, providing 18 data points along the limb and returned the best value $\chi^2_{\text{min}}$= 18.6, which corresponds to a $\chi^2$ per degree of freedom of 1.43, since we have 5 free parameters, which means that the fit was satisfactory.

**UPPER LIMIT ON THE PRESSURE OF A PUTATIVE ATMOSPHERE.** Global atmospheres of the type seen in Pluto and other solar system bodies cause a gradual star disappearance during occultations as well as a gradual reappearance. Upper limits on the pressure of a putative Nitrogen-dominated or Methane-dominated Pluto-like atmosphere can be derived from the occultation profiles, see below. Nitrogen and Methane are the volatile ices that can sublimate at the distances from the sun at which Haumea is (50.593 AU). Even though there is no spectroscopic evidence of $N_2$ or $CH_4$ ices on Haumea's surface in significant amounts[34], their presence below the surface cannot be completely discarded.

Here we use the best data set at hand to derive Haumea's atmospheric upper limits, namely the Asiago light curve, see Fig.1. From Haumea's mass, 4.006 ± 0.04 $10^{21}$ kg (ref. 20), and assuming that the body itself were in hydrostatic equilibrium, we derive an average surface gravity of 0.39 m $s^{-2}$.

We consider first a pure $N_2$ isothermal atmosphere in thermal balance with a surface temperature of 40 K (ref. 16). This provides a vertical density profile from which a ray tracing code calculates synthetic light curves, once a surface pressure $p_{\text{surf}}$ has been prescribed (see refs 4,5 for details). We define a $\chi^2$ function in the same way as in Eq. (1), except that now $p_{\text{surf}}$ is the adjusted parameter. We find that $\chi^2$ is minimum for $p_{\text{surf}}$ = 0 (no atmosphere) and that $\chi^2$ reaches $\chi_{\text{min}}^2$ + 1 for $p_{\text{surf}}$ ($N_2$)= 3 nbar, which sets the 1$\sigma$-level upper limit for a $N_2$ atmosphere. At 3$\sigma$ the upper level is 15 nbar. In Extended Data Fig. 3, we show the effect that a $N_2$ isothermal atmosphere would have on the observations.

A $CH_4$ atmosphere would be more difficult to detect using our data set. This stems from the fact that near-IR heating of methane would cause a thermal profile starting from a typical surface temperature of 40 K, then ramping up to typically 100 K in a few kilometers-thick stratosphere (see discussion in ref. 5). Our data set lacks sufficient spatial resolution to resolve such a thin layer, so that a less constraining upper limit of $p_{surf}$ ($CH_4$)= 10 nbar (1σ-level) can be derived from our observations. At 3σ the upper level is 50 nbar.

In any case, those upper limits are three orders of magnitude below Pluto's atmospheric pressure, meaning that if a global atmosphere exists on Haumea, it is extremely tenuous.

**RING FIT TO THE SECONDARY EVENTS.** Apart from the occultations by the main body, the light curves reveal brief dimmings observed from most of the sites either prior and/or after the main event. The timings of these events were extracted by fitting square-well ring profiles to the short events, in the same way as for Chariklo (ref. 7). However, the only resolved profiles come from the Konkoly 1m telescope (Fig 1). At that station, we derive a radial width (in the ring plane) of $W_{ring}$~74 km at ingress and $W_{\text{ring}}$~44 km at egress, with respective apparent opacities (along the line of sight) of *p'*=0.55 and *p'*=0.56. This implies so-called equivalent widths $W_{\text{equiv}} = W_{\text{ring}} \times p'$ of 41 km and 25 km, respectively, a measure of the radially integrated amount of material contained in the ring. Note that the apparent opacity is related to the apparent optical depth through $\tau'$= -ln(1-*p'*). Converting it into ring normal optical depth $\tau_N$ is not straightforward, as this depends on whether the ring is mono- or polylayer (and it can be neither of them, as these are idealized cases), and is complicated further by diffraction by individual ring particles, see ref. 7.

For the rest of the sites, the ring profiles are not resolved and significant readout times between integration intervals prevent a full recording of each event. So, we used a simple model with a uniform ring of width of 70 km and apparent opacity of 0.5 that provides the typical average equivalent width observed at Konkoly. Those fits account for the readout times between exposures and eventually provide the timings of the synthetic events. Note that in one case (Skalnate egress, Fig 1), the ring is not detected because it should occur during a readout time. Note also that at several stations (Lajatico, San Marcello Pistoiese, Asiago, Wendelstein), the egress ring event is not recorded not by lack of SNR, but because our view of the ring is blocked by Haumea's body (Fig 3).

The locations of the twelve secondary events, projected in the sky plane, allows the retrieval of the full ring orbit, assuming an apparent elliptical shape, and using the same approach as for Haumea's limb fitting, i.e. five adjusted parameters for the ring model. The fit returns a $\chi^2$ per degree of freedom of $\chi^2_{\text{pdf}}$ = 0.43, indicating a satisfactory fit. The radial standard deviation is 27 km, which indicates the typical quality of our fit in a more physical unit. The resulting ring model, where we also outline the assumed physical width of 70 km, is displayed in Fig 3, while the Extended Data Fig. 4 shows an expanded view of the ring northern ansa for a better view.

The parameters of the elliptical fit are $a'_{ring}$= 2,287 $^{+75}_{-45}$ km for the apparent semimajor axis of the ellipse, and $b'_{ring}$= 541 km ± 23 km for the semiminor axis. The Position Angle of the minor axis is $P_{\text{ring}}$= -74.3°±1.3°. Assuming a circular ring, this implies a ring radius $r_{\text{ring}}$= 2,287 $^{+75}_{-45}$ km and opening angle $B_{\text{ring}}$=

asin(*b'/a'*)= 13.8° ± 0.5°. The circular ring assumption is supported by the fact that the center of the ring fitted ellipse and the fitted center of Haumea's limb coincide to within the error bars, see Fig 3. Moreover, the Position Angle $P_{ring}$ = -74.3°±1.3° also coincides with that of the limb minor axis, $P_{limb}$=-76.3°±1.2°. This is another strong argument that we are observing a ring that settled into Haumea's equatorial plane.

The ring radius falls where the 3:1 spin-orbit resonance is expected. That second order resonance occurs when $2\kappa = \omega - n$, where $\kappa$ is the horizontal epicyclic frequency of a particle, $n$ is its mean motion and $\omega$ is Haumea's spin rate. The frequencies $\kappa$ and $n$ are classically given by $n^2=(dU_0/dr)/r$ and $\kappa^2=d(r^4n^2)/r^3$, where $U_0(r)$ is the azimuthally averaged gravitational potential of the primary at distance $r$ from the center. Assuming a homogeneous ellipsoid of semi-axes $a>b>c$ and defining its oblateness as $f= [\sqrt{(a^2+b^2)/2} - c]/a$, then to lowest order in $f$, we obtain $U_0(r) \sim -(GM/r)\ [1 + (f/5)(a/r)^2]$, where $G$ is the gravitational constant, $M$ is Haumea's mass. Using the values *of a,b,c* derived below and $M$= 4.006 ± 0.040 $10^{21}$ kg (ref. 20), we find that the 3:1 resonance occurs at $r_{3:1}$= 2,285±8 km, coincident with the ring radius to within error bars.

**CONSTRAINTS ON THE RING BRIGHTNESS FROM PHOTOMETRIC OBSERVATIONS.** We have modeled the brightness of Haumea and its ring as in refs. 36, 8, 37, to estimate the evolution of the absolute magnitude of the Haumea system as a function of epoch (given that we know the orientation of the ring, the shape of the main body, the geometric albedo of the main body, the radius of the ring, and leaving as free parameter the ring reflectivity I/F, and assuming a width of 70 km for the ring, see previous section). In Extended Data Fig. 5 we plot the absolute magnitude as function of epoch.

To compare the predictions with old observations we used the earliest available magnitude reported to the minor planet center (for 1955), which was based on Digitized Sky Survey (DSS) images of Haumea.The reported value is 16.4 mag in the R band[38]. This can be translated into V magnitude by using the V-R color of Haumea[39], which is 0.33 ± 0.01. Hence, the derived V magnitude was 16.77 which we corrected for phase angle (using the 0.11 mag/degree slope parameter for Haumea[15]) and corrected for heliocentric and geocentric distances to determine an absolute magnitude $H_V$=-0.266 mag for 1955 with an estimated uncertainty of some 20%, not including the rotational variability. For 1991 and 1994 we used the DSS data reported to the minor planet center database, which we assume that can have a photometric uncertainty of at least around 20%. Also, we provided a new data point in 2017 by observing Haumea on several nights using the 1.5m telescope at Sierra Nevada observatory. The standard V and R Johnson Cousins filters were used and Landolt reference stars were observed. This was done in several photometric nights so that an absolute calibration could be performed with good accuracy. The magnitudes were corrected for geocentric and heliocentric distances as well as for phase angle effect using the same coefficient mentioned above. The rotational phase at the time of the observations was accounted for to derive the rotationally averaged absolute magnitude, $H_V$= +0.35 ± 0.06.

The results are plotted in the Extended Data Fig. 5. It turns out that in order to explain the bright absolute magnitude of Haumea in 1955, a ring system with a considerable brightness contribution would seem necessary, but it would provide a too steep behavior that cannot explain the much better data in 2005 and 2017. Hence we can discard it. Our best guess is right now *I/F*~0.09 for a 70 km wide ring. This is comparable to the reflectivity of Chariklo's main ring[7], while being brighter than the similar Uranian rings[40] $\alpha$ and $\beta$, *I/F*~0.05, and dimmer than Saturn's A ring[41], *I/F*~0.5 (all features which have opacities comparable to Haumea's ring). This number should be taken with care, however, as other values of $W_{ring}$ are possible and would provide updated values satisfying $W_{ring} \times I/F \sim$ 7km. Also, considering the large uncertainties in the oldest data, it is clear that more data would be needed to better constrain our model. Note that the photometric observations give support to the pole orientation with J2000 equatorial coordinates (285.1°,-10.6°) rather than (312.3°,-18.6°) because the latter gives a model that is completely out of phase compared to the observations. In particular, it would require an increase of the absolute magnitude from 2005 to 2017, which we can discard. The ring that best fits the photometry data in 2005 and 2017 contributes about ~2.5% of the total flux of the system in 2017.

We also followed the approaches in refs. 8 and 37 to determine the variation of the rotational light curve amplitude with epoch, including a constant contribution from the satellite Hi'iaka. We could reproduce the amplitudes of the rotational variability in 2005, 2007 reported in the literature[1,42] and our own data in 2017 by using a triaxial body with *a*=1,161 km, *b*=852 km and *c*=513 km (see section on Haumea's shape). The ring cannot contribute with more than 5% of the flux because in that case the rate of change of the amplitude is too steep to be compatible with the observations (extended data figure 5, lower panel). Also the pole solution (312.3°,-18.6°) can be discarded as it produces a model that is also out of phase compared to the light curve amplitude observations.

Regarding the ring origin, different mechanisms are possible. Impacts on Haumea might launch material in orbit around it, with some material ending up as a ring inside the Roche limit possibly concentrated in particularly stable regions such as a spin-orbit resonance or near a shepherding satellite. Alternatively, the tidal disruption or collisional disruption of a previously existing satellite could generate a debris disk that would then stay in orbit. Several possible mechanisms of ring formation have been discussed for the case of the ring system found around Chariklo (refs. 7, 43, 44, 45) and Chiron (ref. 8), some of which can also apply to Haumea's case. Also, other scenarios invoking cometary activity as the source of rings[46] cannot apply to Haumea's case, since such an activity is unplausible for a remote and large TNO. In any case, the ring formation process is probably related to the one that formed Hi'iaka and Namaka, and perhaps the orbitally-related Haumea family (ref. 47).

Note finally that some models[48] discuss the possibility that rings can only be found between 8 and 20 AU from the Sun. This is in principle ruled out by the detection of Haumea's ring, currently situated at a heliocentric distance of more than 50 AU, although it cannot be discarded that Haumea could have formed much closer in and migrated to its current position with the ring already formed.

**GEOMETRIC ALBEDO OF HAUMEA.** If the projected size of a solar system body (A) and its brightness are known, we can determine the geometric albedo by using the equation:

$$p_v = \frac{10^{0.4(V_{sun} - H_V)}}{(A/\pi)} \qquad (2)$$

where $V_{sun}$ is the magnitude of the sun ($m_V$=-26.74) and $H_V$ is the instantaneous absolute magnitude. This can be obtained from the rotationally averaged absolute magnitude (<$H_V$>) of Haumea main body. The <$H_V$> determined from the ground includes the contribution of the ring and the satellites Hi'iaka and Namaka, which should be discounted if we want to compute the true geometric albedo of the main body using the occultation effective diameter or projected area. The rotationally averaged <$H_V$> of the Haumea system from ground-based observations is 0.428 ± 0.011 mag (ref. 15). However, that absolute magnitude corresponds to 2005 and we should use the value derived in 2017, plotted in Extended Data Figure 5, which is brighter by 0.07mag. The brightness contribution of the ring to the total brightness of the Haumea system (main body plus ring) was assessed to be almost negligible in a previous section. Hence if we subtract a 2.5% of the brightness to the <$H_V$> measurements of the Haumea system and also subtract a ~11% contribution of Hi'iaka[20,49] and Namaka, the correct $H_V$ value of Haumea's main body at the time of the occultation becomes 0.35 + 0.14 + 0.32/2 (note that we had to add at least 0.32/2 mag because Haumea was at its rotational minimum at the time of the occultation, as explained in the section of the 3-D shape of Haumea). The resulting geometric albedo inserting the area $A$ from the occultation, would be 0.51, which is significantly smaller than the recentmost value of $0.804^{+0.062}_{-0.095}$ derived on the basis of Herschel thermal measurements combined with Spitzer data[11]. Previous modeling of Herschel and Spitzer measurements[16] had resulted in an albedo of 0.7 to 0.75, still significantly higher than the value from the occultation. The main reason for the difference comes from the fact that the occultation results give rise to a larger body than estimated in those works. That might have to do with the beaming parameter or with the phase integral or with deficiencies of the modified thermal models when applied to very elongated TNOs such as Haumea. It must be noted that ref. 50 had already pointed out that thermal models for spherical bodies systematically underestimate the diameters and overestimate the albedos of low-obliquity ellipsoidal asteroids. This seems also applicable to TNOs. Also, the presence of the ring was not anticipated and was not taken into account in the thermal modeling. Hence, Haumea is considerably larger and less reflective than what had been thought before. Thus its surface may contain a larger fraction of rock to ice than estimated in the past. With this value of the geometric albedo, a reanalysis of the spectra of Haumea would imply that the non-icy fraction of the surface can be much larger than the 8% proposed in ref. 51 in which a high geometric albedo was used.

**THREE-DIMENSIONAL SHAPE OF HAUMEA AND DENSITY.** Thanks to rotational light curve measurements performed prior to and after the occultation and given that we know Haumea's rotation period (3.915341 ± 0.000005 h) with a very small uncertainty[16] we could accurately determine the rotational phase at the time of the occultation. It turns out that Haumea was at its absolute brightness minimum, which means that the projected area of the body was at its minimum. This is depicted in Extended Data Fig. 6.

Because the amplitude of the rotational light curve is 0.32 mag (ref. 10), we can use the well known expression that relates the amplitude of the rotational light curve of a triaxial body (ref. 19) in order to derive the three axes of the body:

$$\Delta m = -2.5 log \left( \frac{b}{a} \left( \frac{a^2 \, cos^2(\theta) + c^2 \, sin^2(\theta)}{b^2 \, cos^2(\theta) + c^2 \, sin^2(\theta)} \right)^{1/2} \right) \quad (3)$$

where $a > b > c$ are the semiaxes of the ellipsoid, $\Delta m$ is the amplitude of the light curve (0.32 ± 0.006 mag) and $\theta$ is the so-called "aspect angle", which is the complementary angle to the planetocentric latitude of the subearth point (assumed to be equal to the ring opening angle, as discussed earlier). This aspect angle is thus 90°-13.8°=76.2° ± 0.5° during the occultation in 2017, but $\theta_{09}$=*90°*-6.5=83.5°±0.5° in 2009, when the Hubble Space Telescope observations in ref. 10 were made, using the ring pole position given before. We use the amplitude from ref. 10 because it was determined from Hubble Space Telescope images that could resolve Haumea, Hi'iaka and Namaka so the contribution of the satellites does not affect the photometry, contrary to the case of the ground based observations.

Moreover, using the fact that the the *a*-axis was turned to us during the occultation, we can relate the apparent semi-minor axis in 2017 (*b'*= 569 km) to the true semi-minor axis through $b'^2 = c^2 \sin^2(\theta_{17}) + a^2 \cos^2(\theta_{17})$. Combining this equation to the previous one, and using the numerical values mentioned before, we obtain $a \times b \times c$= (1,161 ± 30)×(852 ± 2)×(513 ± 16) km$^3$. The formal error on *a* mainly comes from the uncertainty on $\Delta m$. The value quoted above (± 0.006) seems too optimistic to represent the oscillation due entirely to shape, so that we used a more plausible value of ±0.02, due to the possible presence of albedo features on Haumea. The formal error on *b* mainly comes from the limb fitting (see Fig. 3), because we are in the special case where the *a*-axis was directed to us during the occultation. Finally, the formal error bar on *c* directly stems from the uncertainty on the apparent semi-minor axis, 569±13 km, that relies again on the limb fitting.

In fact, we note that 0.32 mag is a lower limit because the constant contribution of the ring brightness was not accounted for, so the actual amplitude of the oscillation that would be caused by Haumea alone, without ring, would be larger than 0.32mag. However, the ring contribution is small, as explained in previous sections and the difference in the amplitude of the light curve arising from this is expected to be small. Nevertheless, we cannot discard that unknown satellites or additional rings could also contribute with a few percent.

The density of Haumea, using the value of *a, b, c* derived above and the mass determination (from the orbital period of Hi'iaka[20]) is 1,885 ± 80 kg/m$^3$.For an *a* semi-axis larger than the nominal 1,161 km determined in the present work, the density will be even smaller than 1,885 kg m$^{-3}$. In fact, for a ring brightness of 5% (which is already larger than needed to fit the absolute magnitude data) the longest axis would be 2,520 km, the volume-equivalent diameter becomes 1,632 km and the density would be 1,757 kg $m^{-3}$, which can be considered a lower limit. Values of 1,885 kg $m^{-3}$ to 1,757 kg $m^{-3}$ are far closer to the density of the rest of the large transneptunian objects and is in agreement with the trend of increasing density versus size (e.g. supplementary material in ref. 5, and refs. 21, 22).

Those values can be compared to the limits imposed by the hydrostatic equilibrium of a Jacobi ellipsoid[23]. In this case, the rotational parameter $\omega^2/\pi G\rho$ must be bounded by 0.284 and 0.374 ($\omega$ being the spin frequency of the body and $G$ the gravitational constant). Using a rotation period of 3.915341 h for Haumea, we obtain the condition $2530 < \rho < 3340$ kg $m^{-3}$. This is far from our upper limit 1,885+80= 1,965 kg $m^{-3}$, thus showing that Haumea cannot be a homogeneous ellipsoid in hydrostatic equilibrium.

Note that the axis ratios $b/a$~ 0.73 and $c/a$~ 0.44 are consistent with those of a Jacobi ellipsoid with $\omega^2/\pi G\rho$ ~ 0.35, but the density required by that solution, 2,700 kg $m^{-3}$, is not consistent with our measurement. Ref. 24 had already noted that Haumea's shape might not be that of a fluid equilibrium Jacobi body if granular physics is used to model the shape instead of using the simple assumption of fluid behavior. Ref. 24 also concluded that the density of Haumea could be much smaller than the minimum of 2,600 kg/$m^3$ reported in the literature. For an ellipsoidal body with $b=(a+c)/2$ one can determine an angle of friction between 10° and 15° using figure 4 of ref. 25 for the density of 1,885 kg/$m^3$. A non homogeneous body can also depart from the classical equilibrium shapes. Differentiation can be expected in large bodies such as Haumea, but it remains to be seen whether feasible mass concentrations toward the nucleus can explain the current shape and density.

**ANALYSIS OF THE CRNI VRH IMAGES.** Given that the images were acquired while the telescope was guided at sidereal rate and moving at a speed of 40 arcsec/minute in the North-South direction, with 300s integrations, the resulting images show trailed stars. By using 300s of exposure time and reading out the CCD in 15s the percentage of "deadtime" was only around 5%. This comes at the expense of increasing the background noise considerably. The images were dark-subtracted and flatfielded following standard procedures. To analyze the trail of the occultation star plus Haumea (both blended in a single trail) we took line profiles along the center (and most intense) part of the trail in the different images. Given the known drift of the telescope, we can translate the pixel along the profile to time after the integration. The pixel scale was 1.4 arcsec/pixel. Plots of these line profiles are shown in Extended Data Fig. 7, for images prior to the main occultation and during the main occultation event.

The smoothed line profile of the trail in the image at the time of the occultation shows a clear drop of intensity at an approximate time indicated by the dashed vertical line in Extended Data Fig. 7. This corresponds to the time when the star disappeared due to the occultation. The intensity remains low for the rest of the integration, which means that the star did not reappear while the image was being taken. Hence, from this we can determine a disappearance time and can estimate that the reappearance likely took place while the CCD camera was being readout.

Besides, from the ellipse fit to the occultation chords of the rest of the sites the reappearance time at Crni Vrh is expected to have happened approximately at the time of the readout. The timings derived in this way are those shown in table 1. Even though the time uncertainties are large and difficult to estimate, the data still provided valid information.

**CODE AVAILABILITY.** We have opted not to make our codes available as we cannot guarantee their correct performance under different computing platforms.

**DATA AVAILABILITY STATEMENT.** All relevant data are available from the authors upon request.

## EXTENDED DATA LEGENDS

Extended Data Table 1. Data on the occulted star.

Extended Data Table 2. Timing of the secondary brief occultation events from the different observing sites on 21 January 2017.

Extended Data Figure 1. Declination residuals of Haumea astrometry. Plus symbols: declination residuals of the observed position of Haumea's system photocenter with respect to the theoretical position in declination, from JPL#81 ephemerides. The residuals are shown versus the date of observation. All the observations were obtained with the La Hita 0.77m telescope as explained in Methods. The thin solid line represents a sinusoidal fit to the residuals with the period determined from a periodogram analysis, which is coincident with the orbital period of the moon Hi'iaka. Outlier values have not been removed. In right ascension we did not detect an oscillating behavior of the residuals because Hi'iaka's orbit does not extend as much as in declination and the quality of the data was not good enough to show the periodicity.

Extended Data Figure 2. Map of the Earth showing the locations of the observatories that recorded the occultation (Green dots). The solid lines mark the shadow path limits. Mount Agliale is indicated in blue because the occultation by the main body was not positive there, but the occultation by the ring was detected. The dashed line denotes the center of the shadow path. Note that Munich corresponds to the location of the Bavarian Public Observatory. The complete names of the observatories can be found in table 1. The red marks at Trebur and Valle D'Aosta observatories indicate the two closest sites to the shadow path that recorded a negative occultation. The coordinates of Trebur observatory are 49° 55' 31''.5 N, 8° 24' 40''.6 E and the coordinates of Valle D'Aosta observatory are 45° 47' 22'' N and 7° 28' 42'' E. The shadow motion is from bottom to top of the figure.

Extended data Figure 3. Upper limit on Haumea's $N_2$ atmosphere. The black dots give the normalized flux from the star plus Haumea, as observed from the Asiago

station which is the one that provided the highest SNR and enough time resolution to look for a faint atmosphere. The dots combine ingress and egress data and are plotted against the distance perpendicular to the local Haumea limb, as given by the solution shown in Fig. 2. The horizontal bars associated with each data point indicate the distance interval corresponding to the integration times of each point. The red line shows an example of light curve obtained with an isothermal $N_2$ atmosphere at $T$ = 40 K and with surface pressure $p_{surf}$ = 15 nbar (at 3σ-level upper limit for better illustration, as the 1σ-level of 3 nbar would be difficult to notice). The red open circles show the expected flux at each data point, after convolution with the finite integration point has been performed.

Extended Data Figure 4. Ring upper ansa. An expanded view of Fig. 3 showing in more detail the events along the ring upper ansa. The best fitting mean ring radius $a'_{ring}$ = 2,287 $^{+75}_{-45}$ km is drawn as a solid curve, while the gray area shows the full extension of a semi-transparent 70-km wide ring that is consistent with the twelve secondary events shown in Fig. 3. The lengths of the red segments indicate the uncertainties stemming from the error bars on the ring timings.

Extended Data Figure 5. Photometric models. **a)**, absolute magnitude of Haumea system as a function of time. Diamonds represent observations and lines are models. The continuous thin cyan line represents a model without a ring, the black curve is a model with a ring 70 km wide and with reflectivity *I/F*= 0.09, similar to that of Chariklo's main ring. The ring in this model contributes with approximately 2.5% of the total flux of Haumea plus Hi'iaka in 2017. The blue curve corresponds to a model with a wider (140 km) and brighter (*I/F*=0.36) ring which gives rise to a ~20% contribution of the total brightness in 2017. This can be discarded because it would produce a change in light curve amplitude that is too steep to be compatible with the observations (see lower panel). **b)**, Rotational Light curve amplitude determined from the ground for the same three models of the upper panel (using the same color coding). The diamonds represent observations from the literature and from this work (for 2017). The continuous lines represent the same models as in the upper panel. See Methods for further explanations.

Extended Data Figure 6. Rotational light curve of Haumea. Relative magnitude versus rotational phase obtained 2 days after the occultation with the Valle D'Aosta 0.81m telescope with no filters. The rotational zero phase was established at the time of the occultation and the rotation period used was 3.915341 h. Superimposed is a fit to the observational data. As can be seen, the absolute maximum in magnitude (absolute brightness minimum) is reached at the time of the occultation (arbitrarily located at phase=0 in this plot)), which means that the brightness was at its minimum at the time of the occultation and hence the projected area of Haumea was also at its minimum. The continuous line is a fit to the data. The peak to peak amplitude of the light curve is 0.25±0.02 mag.

Extended Data Figure 7. Profiles of the trails of the occultation star in two images. **a),** profile along a central line in the trail of the occultation star (blended with Haumea) in a drifted image taken from Crni Vrh observatory prior to the main occultation. In ordinates we show the light intensity along the line. Note that the line starts 40 pixels before the beginning of the trail and ends 40 pixels after the trail end. This is done to show the background level and to show that the transition from trail to background is not easy to identify. The horizontal line marks the mean intensity level of the trail. The thick line represents a smoothed profile with a 10-pixel boxcar to filter the high frequency noise. The x-axis has been translated from pixels to time using the drift speed of 40 arcsec/minute, given the known pixel scale of the telescope. The vertical dashed-dotted lines at abscissa 0s and 300s mark the start and end of the integration respectively. The UT at start of exposure was 02:59:19.50. As can be seen, the intensity level is basically constant with time. Note that prior to 0s and after 300s the line profiles decay to the background level of the image because the pixels there were outside the trail. Hence prior to 0s and after 300s the plot does not represent the intensity of the source but the intensity of the background. **b)**, same as upper panel but from the image at the time of the occultation. The UT at the start of the exposure was 03:04:50.11. The dashed vertical line at 185s marks a very approximate moment at which the occultation begins. As can be seen in the smoothed curve, a clear drop of the signal is produced and lasts till the end of the 300s integration.

**EXTENDED DATA TABLES AND FIGURES**

Extended Data Table 1

| | |
|---|---|
| Designation | URAT1 533-182543 GaiaDR1 1233009038221203584 |
| Magnitude[*] | B=20.88, V=17.97, R=18.000, J=15.353, H=14.791, K=14.532 |
| Diameter[**] | ~0.007 mas (~0.27 km at Haumea's distance) |
| Coordinates[***] | $\alpha$ = 14h 12m 03s.2034, $\delta$ = +16° 33' 58".642 |
| Speed[****] | 13.1 km/s |

[*]From the NOMAD catalog
[**]Estimated from the V,B,R,J,H,K magnitudes
[***]Coordinates in the J2000 equinox for epoch 2017.0575 from this work
[****]Speed of Haumea with respect to the star, seen from Earth.

Extended Data Table 2

| Site Name and country | Coordinates Lat dd:mm:ss Lon dd:mm:ss Altitude (m) | Telescope Name Aperture Filter | Detector/Instrument Exposure (s) Cycle time (s) | Mid time of event (UTC) flux standard deviation, level of ring event detection (σ) |
|---|---|---|---|---|
| Skalnate Pleso Observatory -Slovakia | 49°11'21.8" N 20°14'02.1" E 1826 | Astelco 1300/10400 1.3 m no filter | Moravian G4-9000 10 s 15.503 s | Ingress 3:07:04.1±5.6 0.044, 2.1 |
| Skalnate Pleso Observatory -Slovakia | 49°11'21.8" N 20°14'02.1" E 1826 | Astelco 1300/10400 1.3 m no filter | Moravian G4-9000 10 s 15.503 s | Egress* 3:11:50±2.4 0.044, n.a. |
| Konkoly Observatory -Hungary | 47°55'01.6" N 19°53'41.5" E 935 | RCC 1.0 m no filter | Andor iXon 888 1 s 1.007 s | Ingress 3:06:58.1±0.3 0.079, 4.5 |
| Konkoly Observatory -Hungary | 47°55'01.6" N 19°53'41.5" E 935 | RCC 1.0 m no filter | Andor iXon 888 1 s 1.007 s | Egress 3:11:40.7±0.2 0.068, 3.5 |
| Ondrejov Observatory -Czech Republic | 49°54'32.6" N 14°46'53.3" E 526 | Ondrejov 650/2675 0.65 m no filter | Moravian G2-3200 8 s 9.721 s | Ingress 3:06:48.4±4.2 0.041, 2.5 |
| Ondrejov Observatory -Czech Republic | 49°54'32.6" N 14°46'53.3" E 526 | Ondrejov 650/2675 0.65 m no filter | Moravian G2-3200 8 s 9.721 s | Egress 3:10:51.6±3.0 0.041, 2.9 |
| Wendelstein Observatory -Germany | 47°42'13.6" N 12°00'44" E 1838 | 2.0m Fraunhofer f/7.8 2.0 m r' | WWFI** 10 s 14.536 s | Ingress 3:06:39.5±3.9 0.021, 4.9 |
| Asiago observatory Cima Ekar -Italy | 45°50'54.9"N 11°34'08.4"E 1376 | Copernico 1.82 m f/9 1.8m no filter | AFOSC*** 2 s 5.026 s | Ingress 3:06:35.4±0.3 0.029, 8.8 |
| S. Marcello Pistoiese observatory -Italy | 44°03'51.0" N 10°48'14.0" E 965 | 0.60 f/4 0.6m no filter | Apogee Alta U6 10 s 11.877 s | Ingress 3:06:38.8±1.5 0.039, 5.8 |
| Lajatico Astronomical Centre -Italy | 43°25'44.7" N 10°43'01.2" E 433 | RC 50 cm f/9 0.5 m no filter | Moravian G3-1000 15 s 16.254 s | Ingress 3:06:37.9±5.1 0.067, 1.6 |
| Mount Agliale observatory -Italy | 43°59'43.1'' N 10°30'53.8'' E 758 | 50 cm f/4.6 0.5 m no filter | FLI proline 4710 15 s 16.724 s | Ingress 3:06:56.7±1.4 0.031, 4.5 |
| Mount Agliale observatory -Italy | 43°59'43.1'' N 10°30'53.8'' E 758 | 50 cm f/4.6 0.5 m no filter | FLI proline 4710 15 s 16.724 s | Egress 3:08:15.3±4.2 0.031, 5.4 |

*****This egress time is not a real detection of the ring. See Fig. 3.
**http://www.usm.uni-muenchen.de/wendelstein/htdocs/wwfi.html
***http://www.pd.astro.it/index.php/en/asiago-info-eng/who-we-are/136-asiago-eng/250-afosc.html

Extended Data Figure 1

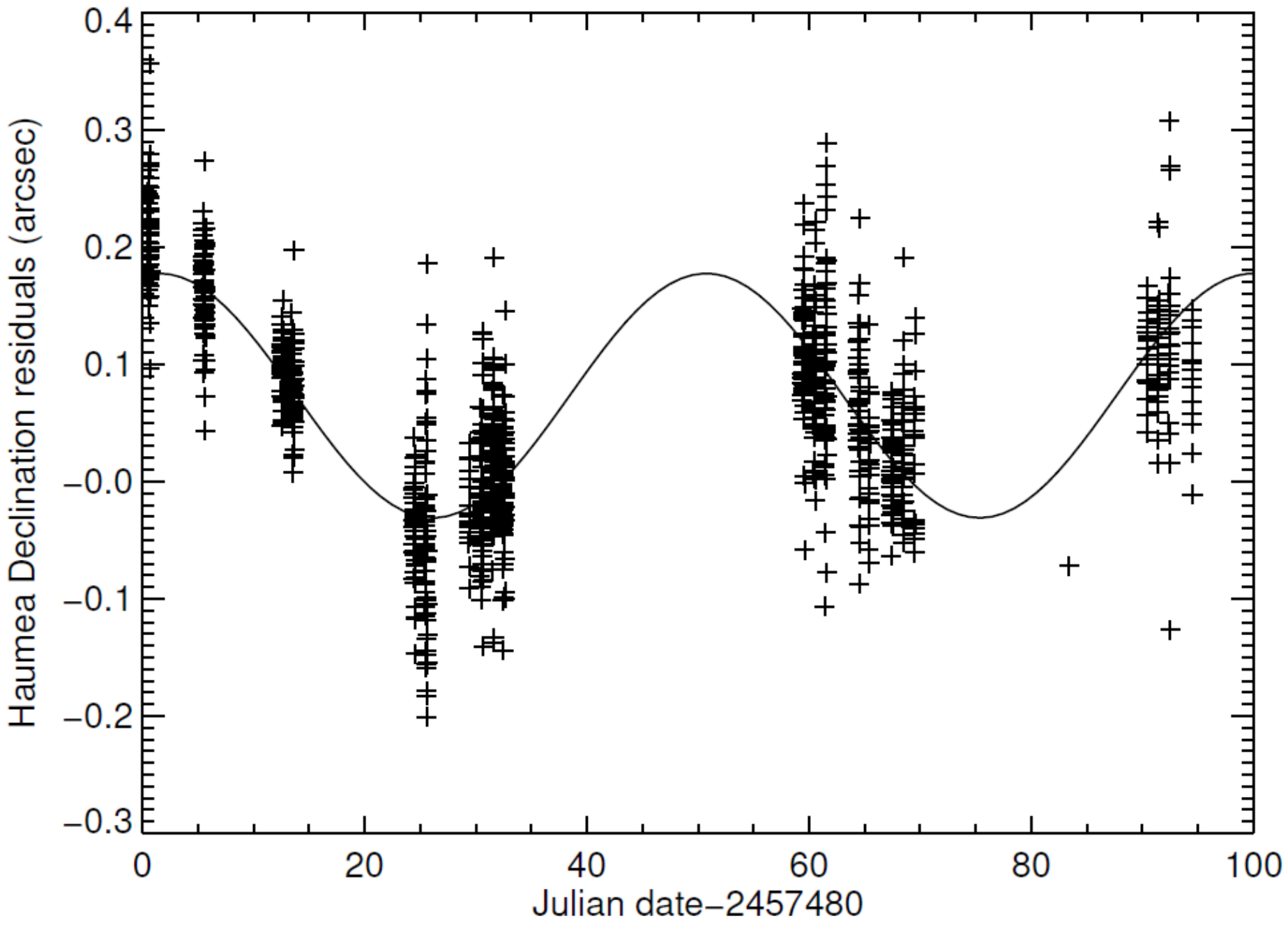

Extended Data Figure 2

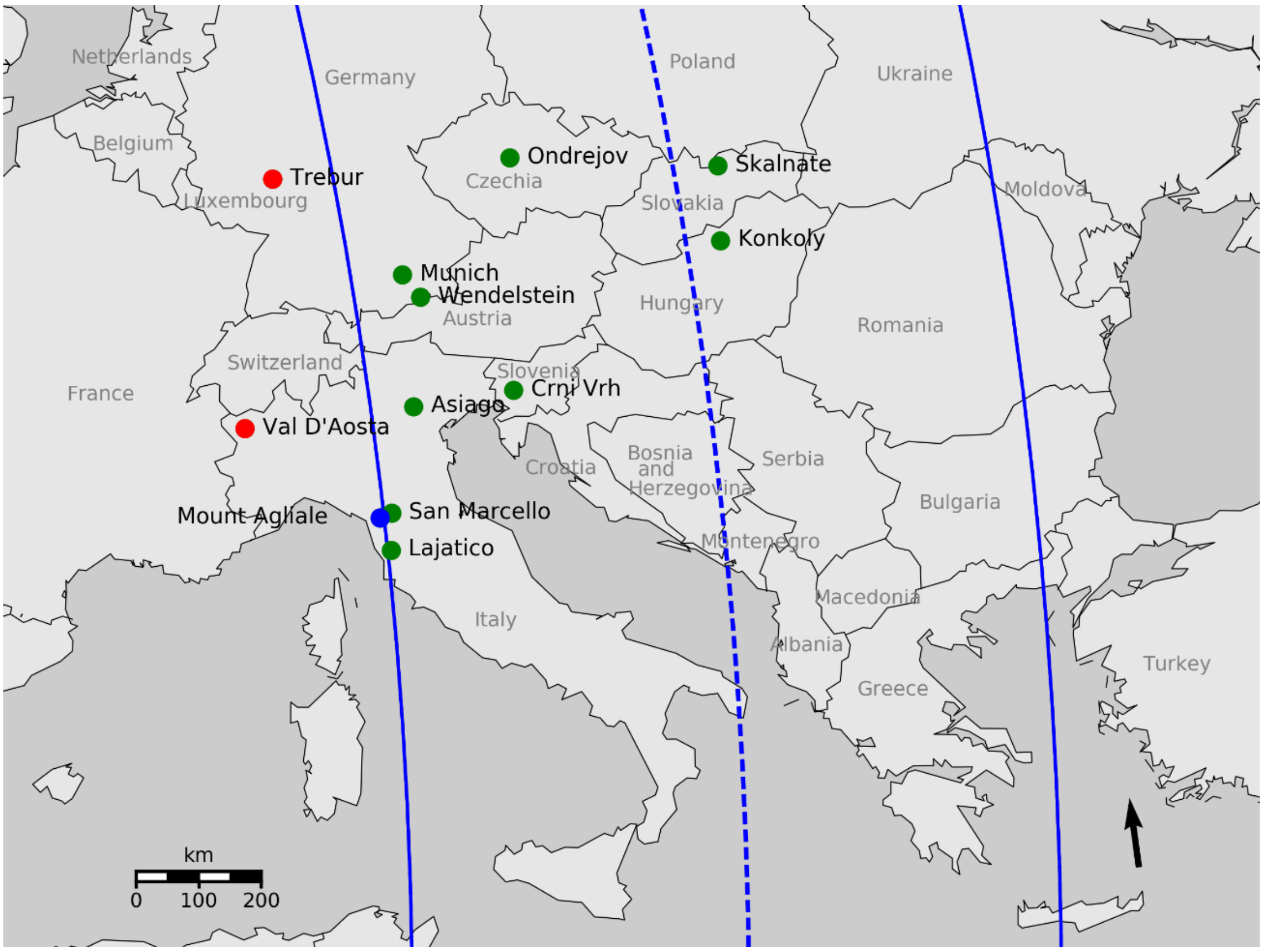

Extended Data Figure 3

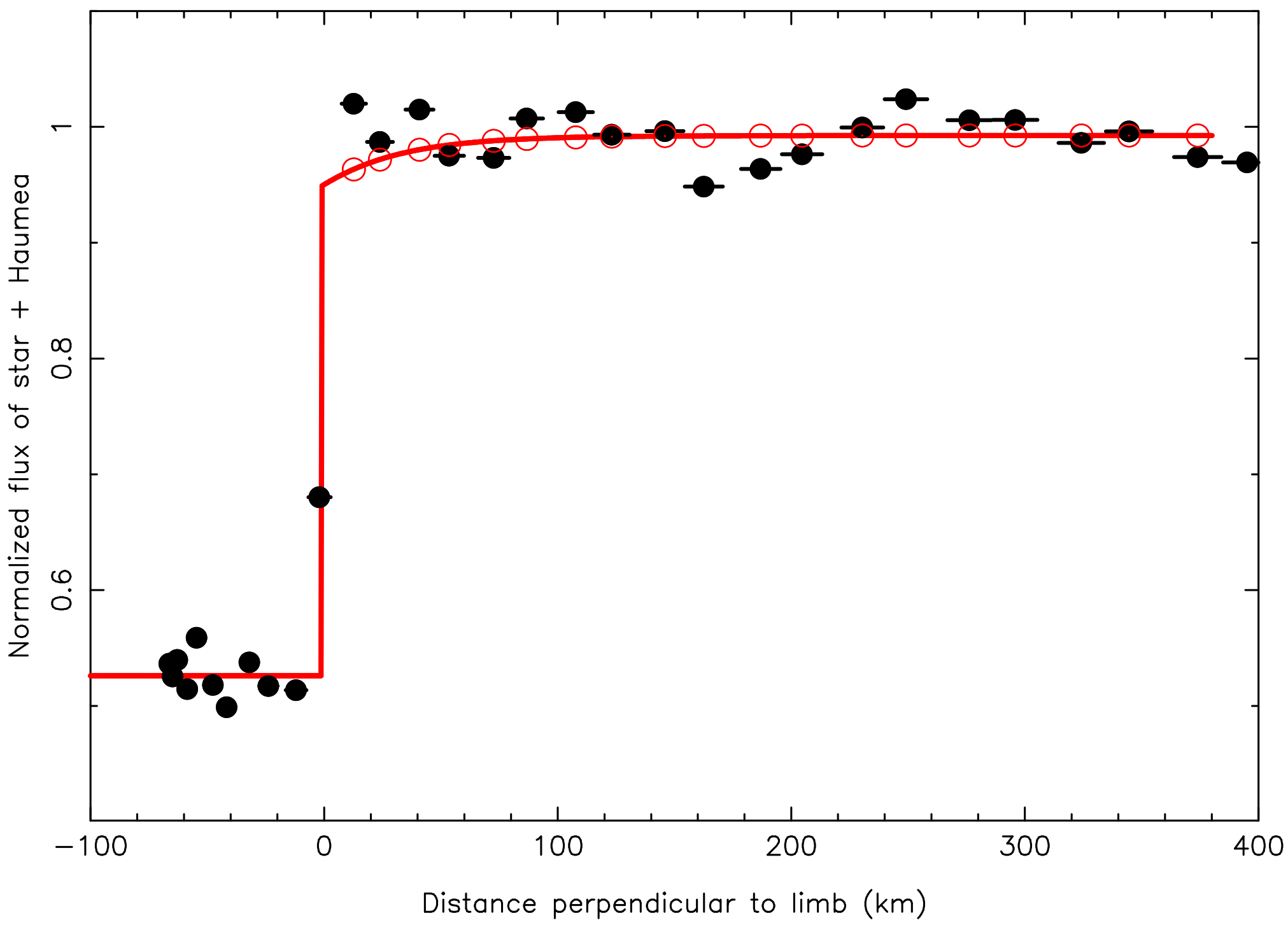

Extended Data Figure 4

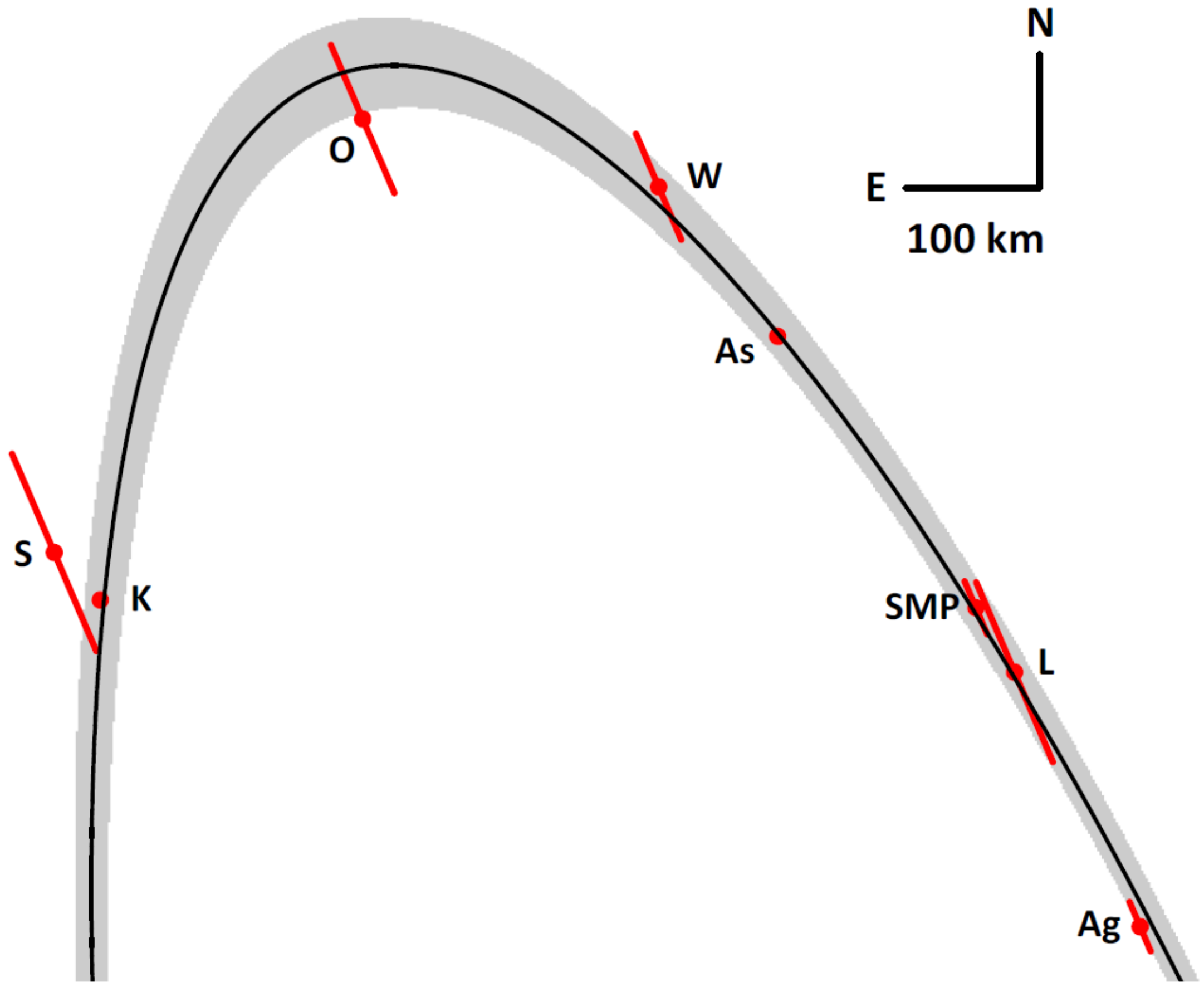

Extended Data Figure 5

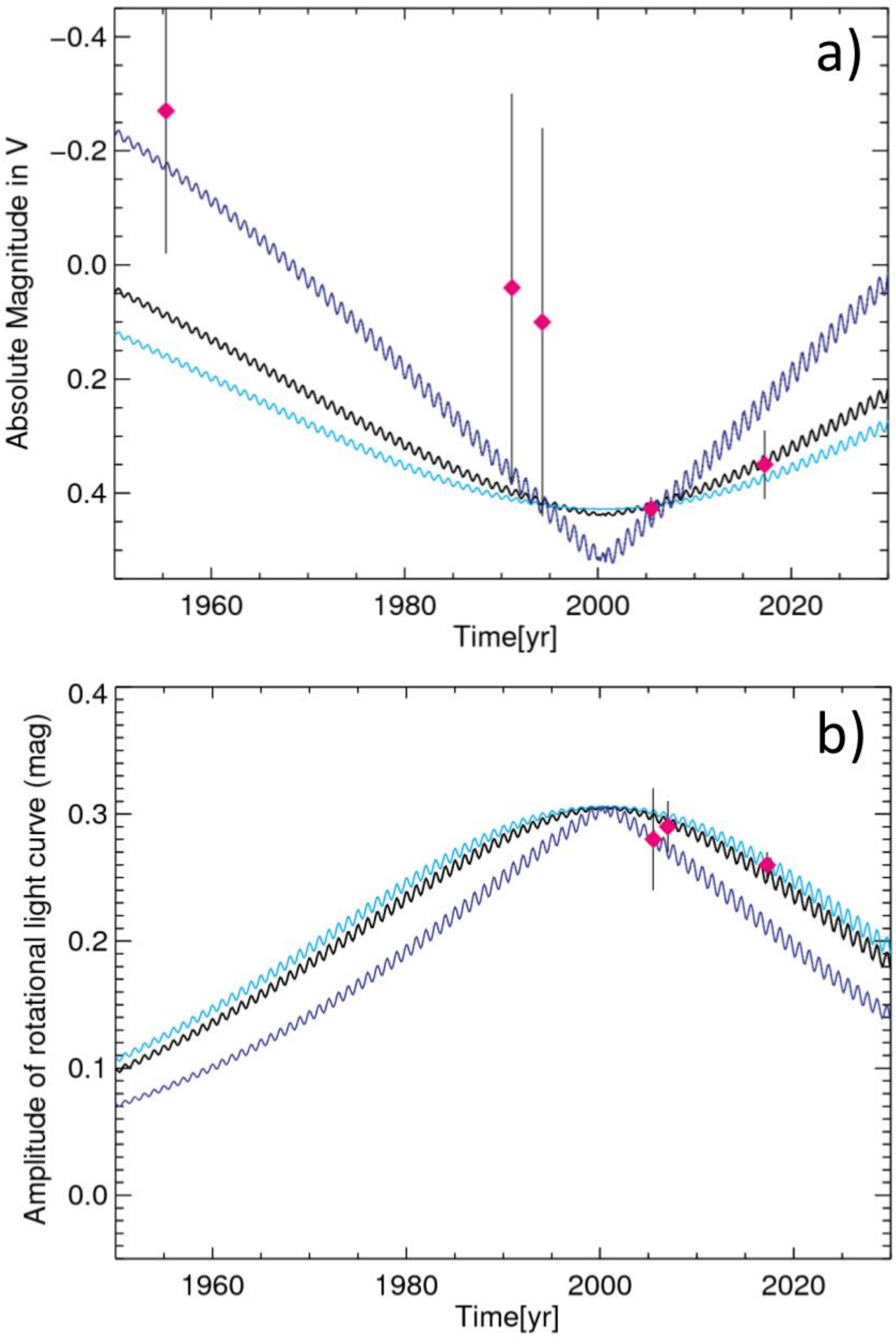

Extended Data Figure 6

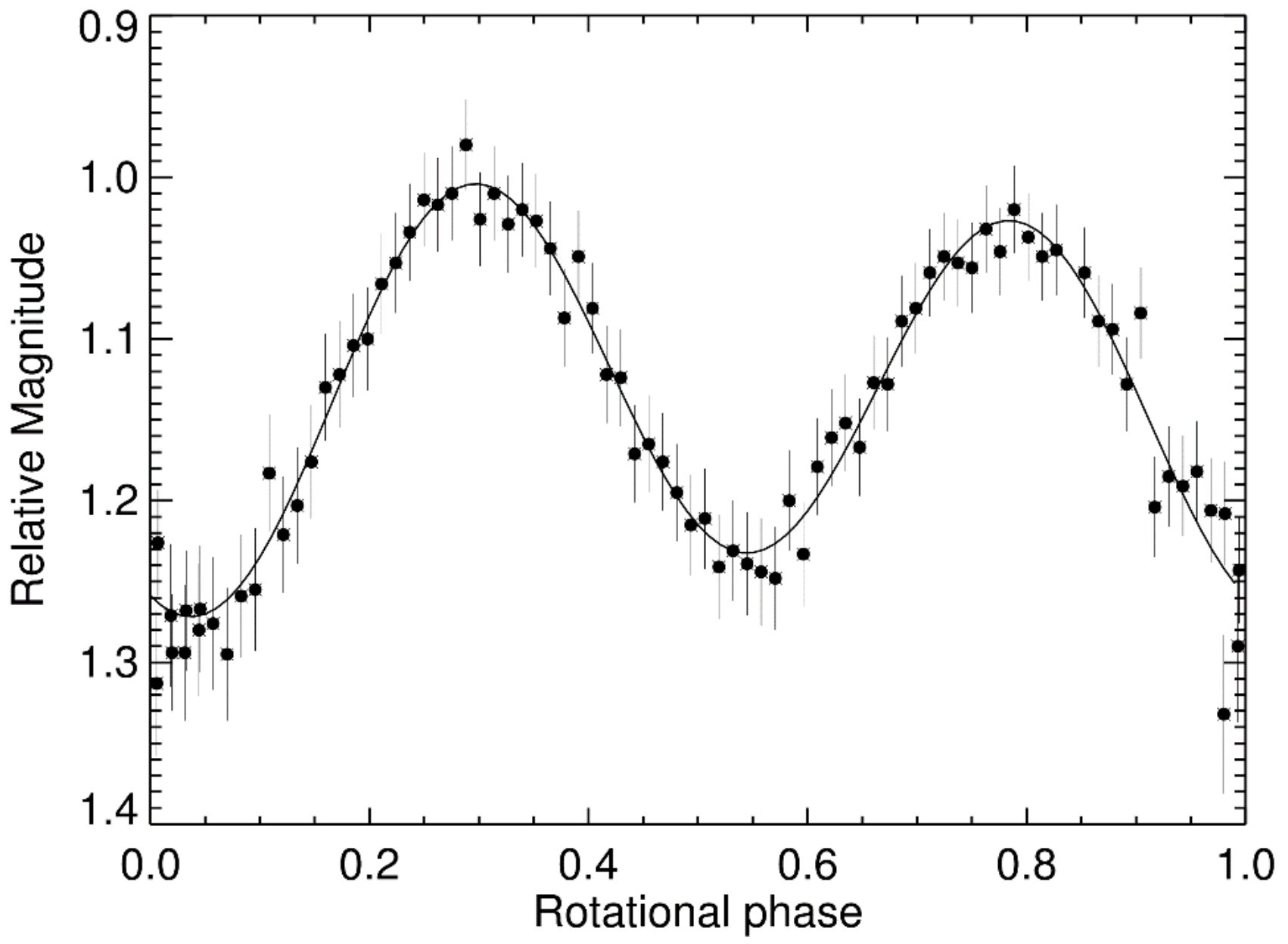

Extended Data Figure 7

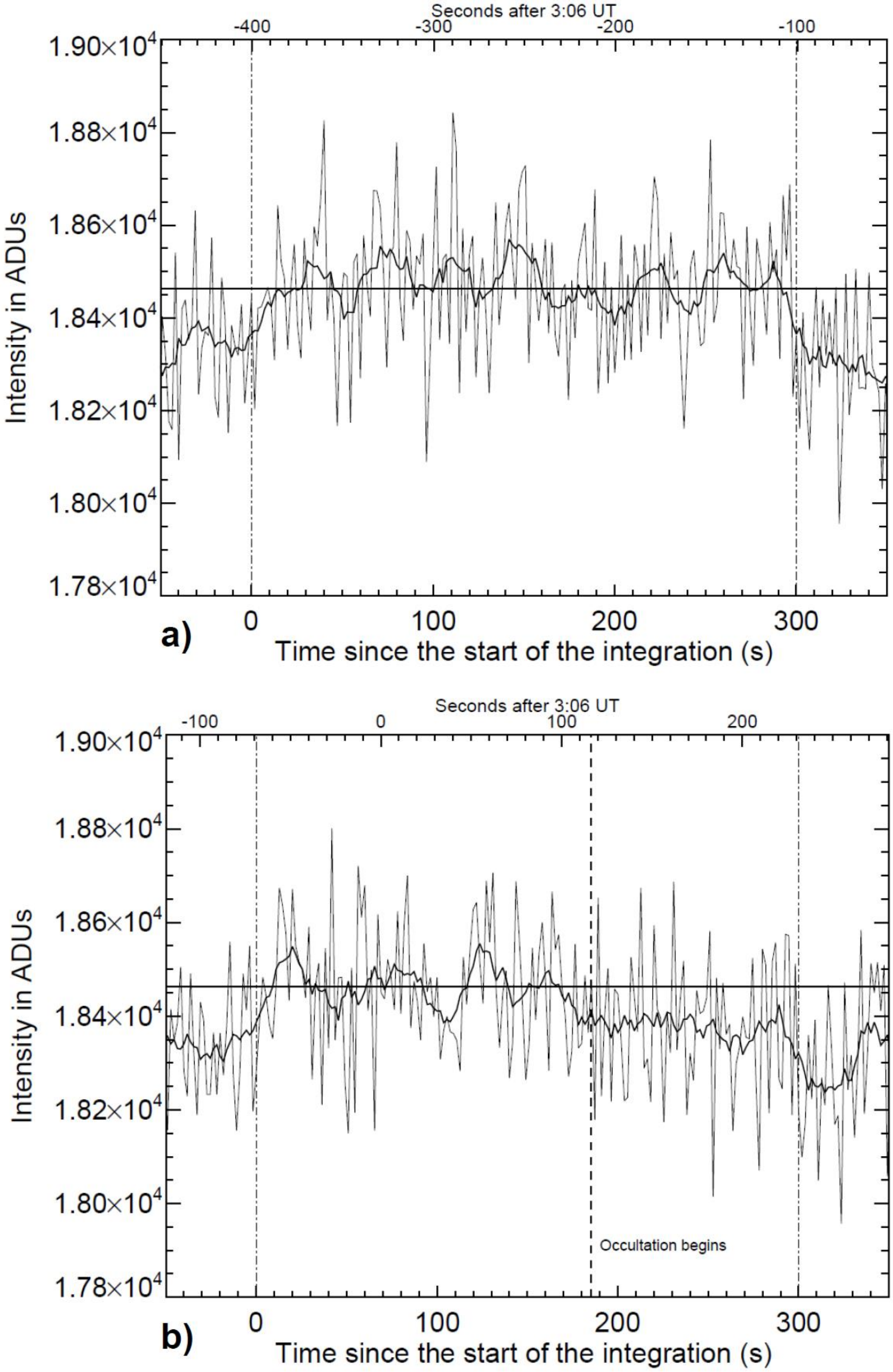